\documentclass[aps,prb,reprint,groupedaddress,floatfix,showpacs]{revtex4-1}
\usepackage{graphicx}
\usepackage{dcolumn}
\usepackage{bm}

\usepackage{color}
\usepackage{ulem}
\begin{document}


%
\title{A first-principles study of carbon-related energy levels in GaN: Part I - Complexes formed by substitutional/interstitial carbons and gallium/nitrogen vacancies}


\author{Masahiko Matsubara}
\author{Enrico Bellotti}
\affiliation{Department of Electrical and Computer Engineering, Boston University, 8 St. Mary's Street, Boston, MA 02215, USA}


\date{\today}

\begin{abstract}
Various forms of carbon based complexes
in GaN are studied with first-principles calculations employing
Heyd-Scuseria-Ernzerhof hybrid functional within the framework of density functional theory.
We consider carbon complexes made of the combinations of single impurities,
i.e.\ $\mathrm{C_N-C_{Ga}}$, $\mathrm{C_I-C_N}$ and $\mathrm{C_I-C_{Ga}}$,
where $\mathrm{C_N}$, $\mathrm{C_{Ga}}$ and $\mathrm{C_I}$ denote C substituting nitrogen,
C substituting gallium and interstitial C, respectively,
and of neighboring gallium/nitrogen vacancies ($\mathrm{V_{Ga}}$/$\mathrm{V_N}$),
i.e.\ $\mathrm{C_N-V_{Ga}}$ and $\mathrm{C_{Ga}-V_N}$.
Formation energies are computed for all these configurations with different charge states 
after full geometry optimizations.
From our calculated formation energies, thermodynamic transition levels are evaluated,
which are related to the thermal activation energies observed in experimental
techniques such as deep level transient spectroscopy.
Furthermore, the lattice relaxation energies (Franck-Condon shift) are computed
to obtain optical activation energies, which are observed in experimental techniques such as
deep level optical spectroscopy.
We compare our calculated values of activation energies with the energies of experimentally
observed C-related trap levels and identify the physical origins of these traps, which were
unknown before.

\end{abstract}


\maketitle

\section{Introduction\label{s:intro}}

Carbon inclusion is unavoidable when growing GaN layers by metal
organic chemical vapor deposition (MOCVD) technique due to several
reasons: metalorganic species used as source materials, contaminants
in the source gases and hydrocarbons from graphite susceptors. As a
result, un-intentional carbon doping is present in GaN layers as
impurities and can assume different configurations in the crystal
lattice. The amount of incorporated carbon depends on the growth
temperature~\cite{koleske02} and pressure~\cite{wickenden04}.
Even in the case of growth performed with molecular beam epitaxy (MBE),
GaN samples are contaminated by carbon impurities as soon as they
are removed from vacuum and various mitigation approaches are being
developed.~\cite{king98,machuca02,koblmuller10}

On the other hand, intentional carbon doping is routinely used to
obtain semi-insulating layers of GaN (GaN:C) that are critical for
the fabrication of high electron mobility transistors
(HEMTs)~\cite{webb99,choi07}. At the same time, the presence of a
significant amount of carbon in the substrate, may lead to deep
level traps acting as recombination centers in the band gap, which
are detrimental to the HEMTs performance, potentially leading to
current collapse and kink effect~\cite{klein99,meneghesso10}.
Therefore in order to understand the effect of carbon on the
devices' operation and improve their performance it is important to
identify the physical origins of C-related deep level traps.

Earlier theoretical calculations based on density functional theory
(DFT) within local density approximation
(LDA)~\cite{boguslawski96,boguslawski97,neugebauer96fest,chisholm01,wright02}
suggested that C could exist as substitutional forms in GaN,
specifically C substituting N ($\mathrm{C_N}$) and C substituting Ga
($\mathrm{C_{Ga}}$). It was also shown that $\mathrm{C_N}$ acts as a
shallow~\cite{shallow}
acceptor with $\sim\,0.3$\,eV activation energy and
C$_{\mathrm{Ga}}$ acts as a shallow donor with activation energy
with $\sim\,0.2$\,eV. It was also suggested that self-compensation
by $\mathrm{C_{Ga}}$ and $\mathrm{C_N}$ pins the Fermi level in the
middle of band gap and this explains the high resistivity of GaN:C
layer~\cite{seager02}. Another possible form of a single
interstitial impurity, C$_{\mathrm{I}}$, was predicted to show
amphoteric behavior, acting as a deep acceptor in $n$-type GaN and
as a deep donor in $p$-type GaN~\cite{wright02}. Furthermore, the
activation energy for this acceptor level due to $\mathrm{C_I}$ has
predicted to be $\sim\,1.1$\,eV.

From the experimental standpoint, a significant number of studies to
understand the trap levels, specifically the energy states in the
band-gap, in carbon doped GaN have been performed using a variety of
techniques. Among these, photoionization spectroscopy
(PS)~\cite{klein99,klein01}, deep level transient spectroscopy
(DLTS), including minority carrier transient spectroscopy (MCTS) and
photoinduced current transient spectroscopy
(PICTS)~\cite{hierro00,armstrong04,armstrong05,fangzq10,honda12,shah12,polyakov13},
deep level optical spectroscopy
(DLOS)~\cite{hierro00,armstrong04,armstrong05,armstrong06} and
cathodoluminescence (CL) measurement~\cite{seager02,polyakov13}.
DLTS, MCTS and PICTS have been primarily employed to detect trap
levels close to the band edges (within $\sim$\,1.0\,eV). Techniques
based on optical methods such as DLOS are used mainly to detect
deeper trap levels, in an energy range close to the center of the
band gap. Combinations of different types of techniques makes it
possible to cover entire band gap region and potentially detect all
existing trap levels.

Using PS technique, Klein \emph{et al.}\ showed that one of two kinds
of deep traps (with absorption threshold at 2.85\,eV), which is
responsible for the current collapse of AlGaN/GaN HEMT, has a carbon
origin because its concentration tracks the carbon doping
level~\cite{klein01}. Hierro \emph{et al.}\ using DLOS measurement
which is able to determine that a trap level located 1.35\,eV below the
conduction band minimum (CBM, $E_c$) was related to
carbon~\cite{hierro00}.
Armstrong \emph{et al.}\ investigated the origin of a number of deep
level traps in GaN using a combination of DLTS, DLOS and
steady-state photocapacitance (SSPC)
techniques~\cite{armstrong04,armstrong05}. Two of the energy levels
obtained by DLTS were ascribed to carbon. One was located 0.11\,eV
below the CBM, and its origin was assigned to $\mathrm{C_{Ga}}$,
based on previous LDA calculated results. The other trap level
appeared at $E_v + 0.9$\,eV, where $E_v$ is the energy of the valence
band maximum (VBM), but its physical origin was unknown. In
addition, four more levels that were obtained by the combination of
DLOS and SSPC were ascribed to carbon. Based on existing LDA
result~\cite{wright02}, a trap energy level at $E_c\,-\,1.35$\,eV
was assigned to an acceptor state of interstitial carbon
($\mathrm{C_I}$). Two traps at $E_c\,-\,1.94$\,eV and
$E_c\,-\,3.0$\,eV were C-related, but their physical forms remained
unknown. A trap at $E_c\,-\,3.28$\,eV was assigned to an acceptor
level of $\mathrm{C_N}$, once again, based on LDA
result~\cite{wright02}.

Shah and coworkers performed DLTS and MCTS
measurements~\cite{shah12} and inferred that an energy level at
$E_c\,-\,0.13$\,eV was possibly related to $\mathrm{C_{Ga}}$ and
behaved as an electron trap state. Furthermore, two energy levels
responsible for trapping holes were observed and were also related to C. One was at
$E_c\,-\,3.20$\,eV with $\mathrm{C_N}$ assignment and the other at
$E_c\,-\,2.69$\,eV with $\mathrm{C_N}$-related defects or gallium
vacancy. Polyakov \emph{et al.}\ studied deep levels by photocurrent
spectra measurements and PICTS. With the former method three
C-related trap levels with optical threshold near 1.3--1.4\,eV,
2.7--2.8\,eV and 3\,eV were observed. The first was attributed to
$\mathrm{C_I}$ acceptor state, whereas the other two were left
unassigned. More recent experimental result, carried out by Honda
\emph{et al.}\ using MCTS measurements, indicated that a hole
trapping state may be present at $E_v\,+\,0.86$\,eV and was assigned
to $\mathrm{C_N}$ based on recent DFT calculation employing hybrid
functionals~\cite{honda12}. They also concluded that one
electron trap at $E_c\,-\,0.40$\,eV was associated with C, but did
not specify its physical form.

%

From the theoretical standpoint, the majority of studies performed
to understand the nature of carbon in GaN have been carried out
using DFT within the LDA
approximation.~\cite{boguslawski96,boguslawski97,chisholm01,wright02}
Only recently, a small number of
investigations~\cite{lyons10,demchenko13,lyons14,alkauskas14} have employed more
sophisticated DFT approaches based on hybrid functionals, with the
intent to overcome the limitations of LDA, and obtain a more
reliable energetics for the various carbon configurations in GaN.
Among these recent studies, DFT calculations using
Heyd-Scuseria-Ernzerhof (HSE) hybrid density functionals for single
carbon impurities, $\mathrm{C_N}$ and $\mathrm{C_{Ga}}$, were
reported~\cite{lyons10,demchenko13,lyons14}. Unlike in previous LDA results,
these calculations indicates that $\mathrm{C_N}$ may not be a
shallow but a deep acceptor with $\sim 0.9$\,eV activation energy,
which means that C cannot be used as a $p$-type donor in GaN.
Furthermore, recent HSE-based calculations, indicates that the
acceptor level of $\mathrm{C_I}$ is $\sim 0.4$\,eV as opposed to
the values of $\sim 1.1$\,eV obtained using LDA.~\cite{lyons14}.
These new outcomes, suggest that the assignments of experimentally
observed trap levels based on earlier LDA results should be
re-examined using more reliable approaches, such as HSE hybrid
density functionals, that could provide a more accurate picture of
the system energetics.

The aim of this work is twofold, first, we intend to perform a
comprehensive study of the formation energies of single carbon
impurity and complexes in GaN using state-of-the-art DFT and HSE
hybrid density functionals. Specifically we intend to focus on the
role of complexes about which very little is known, although
$\mathrm{C_N}$ is considered as the dominant form as a single carbon impurity
especially in $n$-type GaN in Ga-rich conditions.
Indeed, the origins of some of the experimentally observed C-related
trap levels are assigned to $\mathrm{C_N}$ as described above.
Second, we
intend to perform a systematic comparison of the numerical results
with the available experimental data with two specific goals in
mind: establish which energy level can be reliably assigned to a
given configuration, and for which energy level additional
experimental and theoretical work is needed. We want to emphasize
this last issue, since there are several experimentally observed
C-related energy levels, whose physical forms are still unknown.

This manuscript focuses on the complexes that carbon forms with Ga,
N and their vacancies. Furthermore, this work is a companion to a
second manuscript\cite{Matsubara_C2} in which we present the
investigation of complexes that carbon forms in GaN with silicon,
oxygen and hydrogen and we discuss their relative concentrations.

This paper is organized as follows. In Section~\ref{s:methods}, the
details of the computational model are presented and the theoretical
formulations of the formation energies and transition levels are
provided. Furthermore we describe what kind of convergence studies
have been performed to determine the supercell size to be used in
the case of charged defects and complexes. In
Section~\ref{s:results}, we will outline our calculated results that
will be discussed in Section~\ref{s:discussion}. Finally concluding
remarks are given in Section~\ref{s:conclusion}.

\section{Methods\label{s:methods}}

In this section the computational approach is outlined. First the
computational framework is presented. Subsequently the model used to
evaluate the formation energy is given, and finally the
effect of the supercell size on formation energy is discussed.

\subsection{Computational Approach\label{ss:comp}}

\begin{figure}
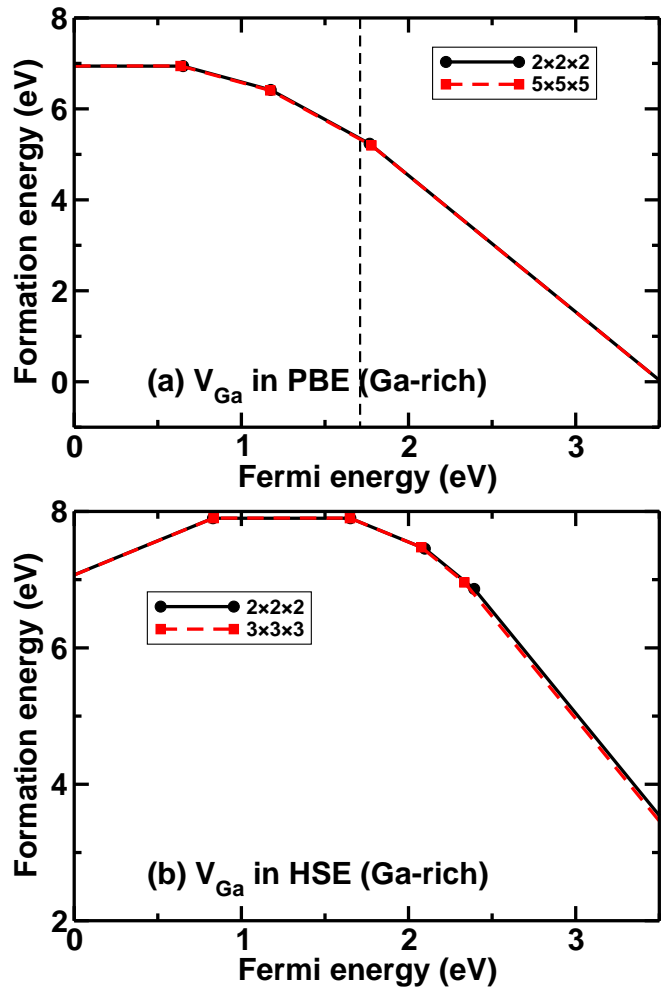

\includegraphics[width=\columnwidth]{fVGa-Garich-kconvPBE-CBM.eps} \\

\includegraphics[width=\columnwidth]{fVGa-Garich-kconvHSE.eps}
\caption{\label{f:kconv}Formation energies for Ga vacancy with different
sets of $k$-point meshes in (a) PBE and (b) HSE. In the case of PBE, the
$2\times2\times2$ and $5\times5\times5$ meshes are compared, whereas in the case of HSE
$2\times2\times2$ and $3\times3\times3$ meshes are compared. Note that no band gap
correction was done in the case of PBE (vertical dashed line at 1.71\,eV denotes the
calculated conduction band minimum).}
\end{figure}

The calculations presented in this work were carried out  using the
projector augmented wave (PAW) method~\cite{blochl94} implemented in
the VASP code~\cite{kresse96,*kresse99}. The main results were
obtained using Heyd-Scuseria-Ernzerhof (HSE) hybrid
functionals~\cite{heyd03,*heyd06}. Additionally, convergence tests
were also performed using standard Perdew-Burke-Ernzerhof
(PBE)~\cite{perdew96} functionals. The semicore Ga $3d$ electrons
were included as valence, since treating Ga $3d$ electrons as core caused relatively large
errors ($\sim 0.5$\,eV) in the formation energies for carbon
complexes. Furthermore, in the case of HSE, in order to reproduce
experimental band gap value (3.5\,eV~\cite{monemar74}), the amount of
exact exchange was taken to be 28\% (giving 3.45\,eV band gap value).
Spin is explicitly considered (spin-unrestricted) in all the calculations.
Finally, a 425\,eV cutoff energy
was used. During the structural optimization procedure to obtain
the total energy of different configurations, the atomic positions
were allowed to change until the largest force component was less
than 0.05\,eV/\AA. The bulk parameters obtained from fully optimized
unit cell are summarized in Table~\ref{tab:bulkpara}. Using these
optimized lattice constants, a supercell containing total 96 atoms
with orthorhombic shape was constructed and employed to study the
carbon inclusions.
The Brillouin zone is sampled on a mesh composed
of a $2\times2\times2$ $k$-point grid. Convergence tests in PBE showed that
the differences in formation energies between $2\times2\times2$ mesh
and denser $5\times5\times5$ mesh were less than 50\,meV. The same tests
were also done in HSE, where $2\times2\times2$ mesh and $3\times3\times3$ mesh
were compared. The differences in formation energies were less than 0.1\,eV.
In Fig.~\ref{f:kconv}, the formation energies (its definition will be provided
in the next section) for $\mathrm{V_{Ga}}$ in Ga-rich condition are presented
for different $k$-point mesh both in PBE and HSE to show the convergence.
Convergence tests performed to evaluate the dependence of the calculated values
on the supercell size will be discussed in Section\,\ref{s:supcell}.


\begin{table}
\caption{Calculated values for lattice constants, energy gap, formation enthalpy and static dielectric constant. 
Experimental results are also given.\label{tab:bulkpara}}
\begin{ruledtabular}
\begin{tabular}{lllllll}
       & $a$ (\AA) & $c$ (\AA) & $u$ & $E_{\mathrm{gap}}$ (eV) & $\Delta H_f$ (eV) & $\varepsilon_0$\footnotemark[1] \\ \hline
Calc. & 3.178 & 5.171 & 0.377 & 3.45 & $-1.18$ & 9.48 \\
Expt.  & 3.190\footnotemark[2] & 5.189\footnotemark[2] & 0.375\footnotemark[2] & 3.5\footnotemark[3] & $-1.63$\footnotemark[4]/$-1.14$\footnotemark[5] & 9.8\footnotemark[6]   \\
\end{tabular}
\end{ruledtabular}
\footnotetext[1]{averaged value over $E_{\parallel c}$ and $E_{\perp c}$ components.}
\footnotetext[2]{Ref.~\onlinecite{xuyn93}.}
\footnotetext[3]{Ref.~\onlinecite{monemar74}.}
\footnotetext[4]{Ref.~\onlinecite{ranade00}.}
\footnotetext[5]{Ref.~\onlinecite{hahn40}.}
\footnotetext[6]{Ref.~\onlinecite{barker73}.}
\end{table}

\subsection{Defect Formation Energies and Thermodynamic Transition Levels\label{ss:Ef}}

The formation energy ($E_f^q$) as a function of the Fermi energy
($E_F$), for a given defect configuration $D$ was calculated with
the following formula :

\begin{eqnarray}
E_f^q (D, E_F) = E_{\mathrm{tot}}^q(D) &-& E_{\mathrm{bulk}} - \sum_{\mathrm{X}} n_{\mathrm{X}}\mu_{\mathrm{X}} \nonumber \\
&+& q (E_F + E_v) + \Delta E^q_{\mathrm{corr}}\label{eq:Ef},
\end{eqnarray}

where $E_{\mathrm{tot}}^q(D)$ is the total energy of the system with
a defect $D$ in a
charge state $q$, $E_{\mathrm{bulk}}$ the total energy of bulk
wurtzite GaN, $n_{\mathrm{X}}$ the number of X (X = Ga, N or C)
atoms removed from, or added to, the system with the chemical
potential $\mu_{\mathrm{X}}$ and $E_v$ the energy of the valence
band maximum (VBM). The last term is the correction for charged
defects in the finite supercell. In this work, we have adopted the
method proposed by Freysoldt \emph{et al.}\ to calculate this
correction term~\cite{freysoldt09,*freysoldt11} using sxdefectalign
program~\cite{sxdefectalign}.


The chemical potential for Ga ($\mu_{\mathrm{Ga}}$) was evaluated
using bulk $\alpha$-Ga and that for N ($\mu_{\mathrm{N}}$)
determined using an isolated N$_2$ molecule. The value of
$\mu_{\mathrm{C}}$ was obtained from the calculated value for cubic
diamond. Furthermore, $\mu_{\mathrm{Ga}}$ and $\mu_{\mathrm{N}}$
satisfy the following condition
\begin{eqnarray}
\mu_{\mathrm{Ga}}+\mu_{\mathrm{N}}=E(\mathrm{Ga})+\frac{1}{2}E(\mathrm{N_2})+\Delta H_f(\mathrm{GaN}),\label{eq:HfGaN}
\end{eqnarray}
where $\Delta H_f(\mathrm{GaN})$ is the formation enthalpy of GaN. In the
Ga-rich limit, $\mu_{\mathrm{Ga}}$ corresponds to the energy of bulk
Ga ($E(\mathrm{Ga})$), whereas in the N-rich limit,
$\mu_{\mathrm{N}}$ corresponds to the half value of the energy of $\mathrm{N_2}$ ($\frac{1}{2}E(\mathrm{N}_2)$).
The thermodynamic transition energy is defined
as the position of Fermi level at which the most stable charge state
changes from $q$ to $q^{\prime}$:
\begin{eqnarray}
\epsilon(q/q^{\prime}) = \frac{E_f^q(D, E_F=0)-E_f^{q^{\prime}}(D, E_F=0)}{q^{\prime}-q}.\label{eq:eqq}
\end{eqnarray}
Since this formulation of formation energy is based on thermodynamic
equilibrium, the calculated transition level is directly related to
the thermal activation energies obtained by thermal experimental
technique such as DLTS. In addition, under thermodynamic equilibrium
condition, the concentration, [C], of an impurity with the formation
energy $E_f$ can be computed by the equation
\begin{eqnarray}
[\mathrm{C}]=N\exp\left(\frac{-E_f}{k_BT}\right),\label{eq:conc}
\end{eqnarray}
where $N$ is the number of defect sites per volume, $k_B$ is the
Boltzmann's constant and $T$ is the growth temperature. As a result
we expect that, the lower the formation energy is, the higher is the
concentration of a specific defect configuration.

\begin{figure}
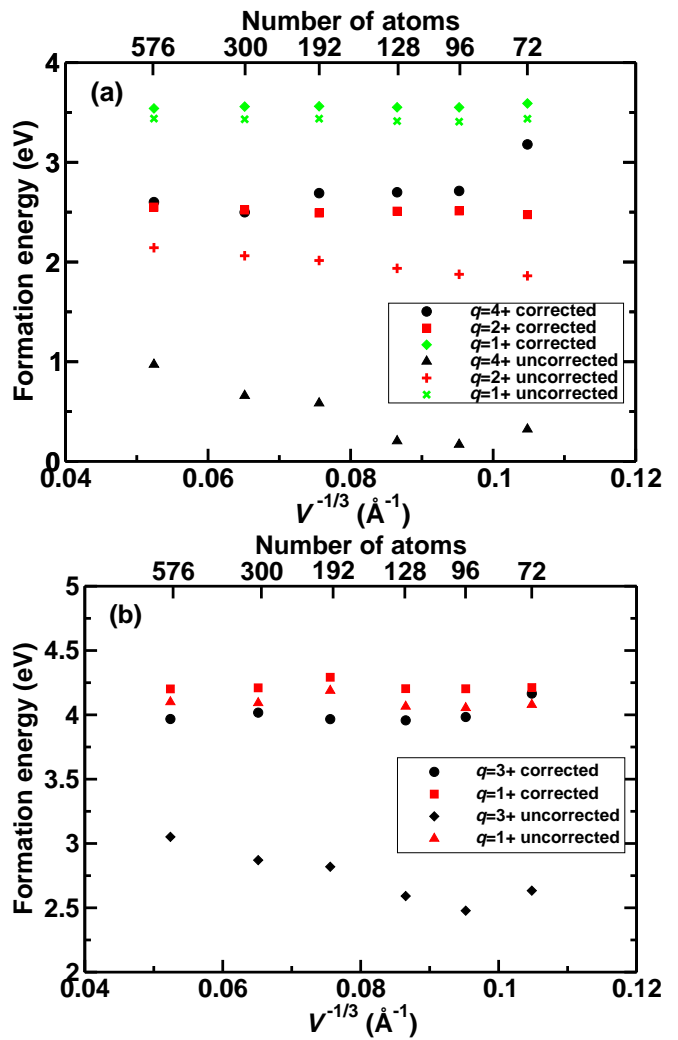

\includegraphics[width=\columnwidth]{CIcorr.eps} \\

\includegraphics[width=\columnwidth]{CICNcorr.eps}
\caption{\label{f:conv}Formation energies with respect to the
supercell size for (a) $\mathrm{C_I}$ with $4+$, $2+$ and $1+$
charge states and (b) $\mathrm{C_I}-\mathrm{C_N}$ complex with $3+$,
and $1+$ charge states. Both corrected and uncorrected formation
energies are plotted for comparison.}
\end{figure}

\subsection{Supercell size for charged defects}
\label{s:supcell}

Defect formation energies calculations are customarily performed
within a periodic supercell approach. However, this approach, when
applied to charged systems, is hampered by spurious Coulomb
interactions between defect itself and its periodic images. In order
to exclude this artificial effect, a number of correction schemes
have been
proposed~\cite{makov95,lany08,lany09,freysoldt09,freysoldt11,taylor11}.
For this work we adopted the scheme proposed by Freysoldt \emph{et
al.}~\cite{freysoldt09,freysoldt11}. We performed a series of test
calculations for a number of significant defect configurations to
check the convergence of the formation energies with respect to the
supercell size (72-, 96- 128-, 192-, 300-, 576-atom supercells). For
these convergence test calculations, we used Perdew-Burke-Ernzerhof
(PBE)~\cite{perdew96} functionals, since convergence studies using HSE are
more involved due to the computational demand of this approach.
However, we can expect that if the convergence is reached in PBE,
the same is true in HSE, because the correction scheme works better in
hybrid functional calculations~\cite{komsa12}.
Convergence test results (both corrected
and uncorrected) are shown in Fig.~\ref{f:conv} for $\mathrm{C_I}$
with $4+$, $2+$ and $1+$ charge states and for
$\mathrm{C_I}-\mathrm{C_N}$ with $3+$ and $1+$ charge states.
In the case of $\mathrm{C_I}$, convergence is already reached with a
72-atoms supercell both for corrected and uncorrected results with
the $1+$ charge state. For the $2+$ charge state, differences
between corrected and uncorrected results are noticeable
particularly in smaller size supercells. Nevertheless, the corrected
formation energy is already well converged at the 72-atom supercell.
Finally, for the $4+$ charge state the differences between corrected
and uncorrected results are sizable even at the 576-atom supercell.
However, for this charge state, convergence is reached at the
96-atom supercell after the correction. In the case of complexes,
such as $\mathrm{C_I}-\mathrm{C_N}$, the situation is almost the
same as in the case of $\mathrm{C_I}$. Specifically, for the $1+$
charge state, the difference between corrected and uncorrected
results are small and convergence is reached with the 72-atom
supercell. For the $3+$ charge state, differences between corrected
and uncorrected results are large and the correction is significant.
In this charge state, 96-atom supercell gives converged results
after the correction in applied.
Based on the outcome of these convergence studies, for our
computation we adopted 96-atom supercell and between two and six
integration points, to obtain accurate results with reasonable
computational costs.


\section{Results\label{s:results}}

In this section we will briefly present the results obtained for
single carbon impurities. Subsequently, we will discuss in details
the outcome of the calculations for carbon complexes that have not
been as extensively studied as the single carbon impurities.

\subsection{Single carbon impurity\label{ss:singleC}}

\begin{figure}
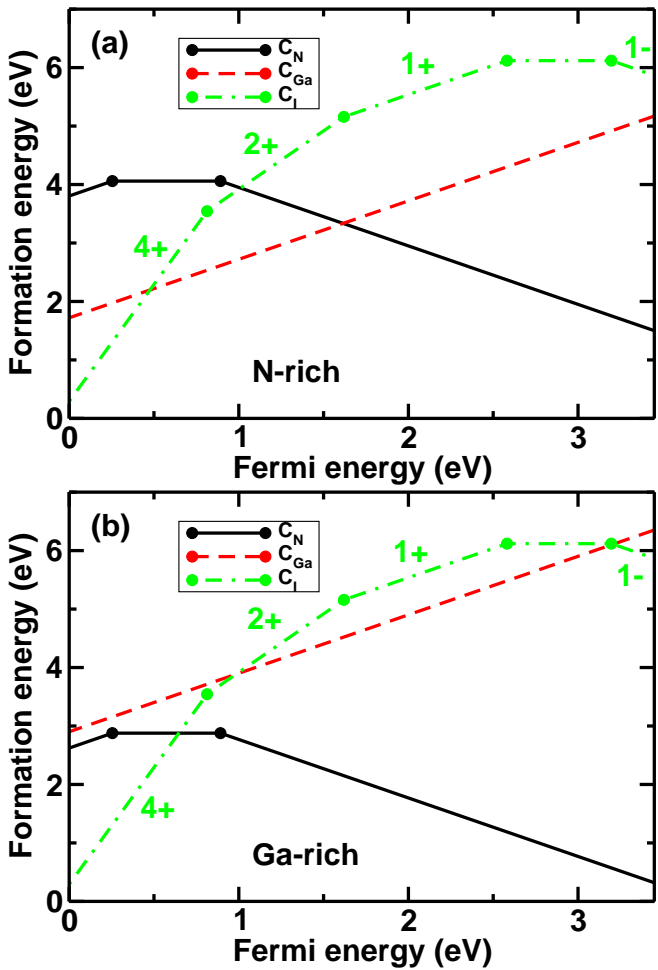

\includegraphics[width=\columnwidth]{fC-Nrich.eps} \\
\includegraphics[width=\columnwidth]{fC-Garich.eps}
\caption{\label{f:EfC}Formation energies as a function of Fermi energy for C$_{\mathrm{N}}$ (solid line in black), C$_{\mathrm{Ga}}$ (dashed line in red) and C$_{\mathrm{I}}$ (dashed-dotted line in green) in (a) N-rich conditions and (b) Ga-rich conditions.}
\end{figure}

This section presents the results obtained using HSE for single
carbon, substitutional and interstitial, impurities:
$\mathrm{C_{Ga}}$, $\mathrm{C_N}$ and $\mathrm{C_I}$.
The calculated formation energies for these defects are shown in
Fig.~\ref{f:EfC}, where only the lowest energy states within the
band gap are presented.
Among them, our emphasis is on the $\mathrm{C_I}$, because two substitutional
cases, $\mathrm{C_{Ga}}$ and $\mathrm{C_N}$, have been studied in detail both within
LDA~\cite{boguslawski96,boguslawski97,neugebauer96fest,chisholm01,wright02} and HSE~\cite{lyons10,demchenko13,lyons14}. For $\mathrm{C_{Ga}}$ and $\mathrm{C_N}$ we
provide a summary of the structural and electronic properties in Table~\ref{t:CGaCN} together
with the values from literature for comparison.
\begin{table*}
\caption{Comparison of the calculated structural and electronic properties of $\mathrm{C_N}$ and $\mathrm{C_{Ga}}$ in the different charge states $q$. The averaged bond change ($\Delta l$ given in \%) is defined as the change of C--N (C--Ga) bond lengths in $\mathrm{C_{Ga}}$ ($\mathrm{C_N}$) against the bulk Ga--N bond values. Negative (positive) values describe the decrease (increase) of the bond lengths.
The formation energies ($E_f$ given in eV) in Ga-rich conditions are also presented.\label{t:CGaCN}}
\begin{ruledtabular}
\begin{tabular}{cccccccccc}
 & & \multicolumn{2}{c}{LDA~\footnotemark[1]} & \multicolumn{2}{c}{LDA~\footnotemark[2]} & \multicolumn{2}{c}{HSE~\footnotemark[3]} & \multicolumn{2}{c}{HSE (this work)} \\ \hline
Form & $q$ & $\Delta l$ & $E_f$ & $\Delta l$ & $E_f$ & $\Delta l$ & $E_f$ & $\Delta l$ & $E_f$ \\
$\mathrm{C_{Ga}}$ & $1+$ & $-17.3$ & -- & $-19$ & $3.17+E_F$ & $-26$ & $2.7+E_F$ & $-19.5$ & $2.90+E_F$ \\
                  & $0$ & $-18.1$ & 5.7 & -- & 6.45 & -- & -- & -- & -- \\
$\mathrm{C_N}$ & $1+$ & -- & -- & -- & -- & +6.7 & $2.6+E_F$ & +6.1 & $2.62+E_F$ \\
               & $0$ & $-2.0$ & 1.1 & $+0.3$ & 2.62 & +2.75 & 2.9 & +2.2 & 2.88 \\
               & $1-$ & -- & -- & $-1.2$ & $2.88-E_F$ & $-2$ & $3.8-E_F$ & $-1.0$ & $3.77-E_F$ \\
\end{tabular}
\end{ruledtabular}
\footnotetext[1]{Refs.~\onlinecite{boguslawski96,boguslawski97}.}
\footnotetext[2]{Ref.~\onlinecite{wright02}.}
\footnotetext[3]{Refs.~\onlinecite{lyons10,lyons14}.}
\end{table*}
%

\begin{figure}
\includegraphics[width=\columnwidth]{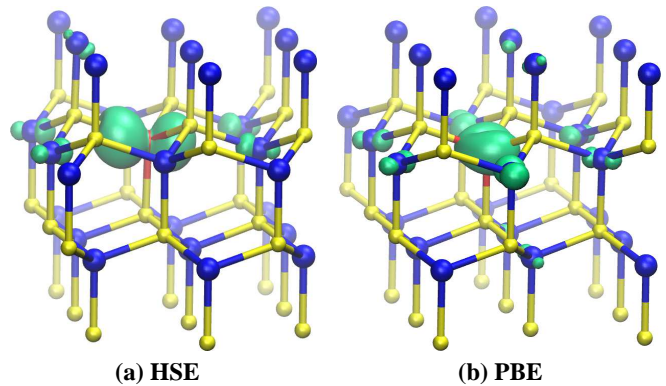}%
\caption{\label{f:CN0spinden}Spin density of the $\mathrm{C_N}$ in the 0 (neutral)
charge state obtained within (a) HSE and (b) PBE.
Isosurface values are taken to be 0.05 $\mu_B$/\AA$^3$.}%
\end{figure}

In the cases of $\mathrm{C_N}$ with $q=0$ and $q=1+$, their structures show
large symmetry lowering due to the Jahn-Teller distortion, which are not seen
in the standard LDA/GGA calculations~\cite{lany10}. With the $q=0$ case as an example,
after the PBE relaxation, $\mathrm{C_N}$ occupies the high symmetric position,
where the distance between N atom parallel to the $c$ axis (1.96 \AA) and those between
N atoms perpendicular to the $c$ axis (1.97 \AA) are nearly the same.
On the other hand, after the HSE relaxation, C--N distance perpendicular to the $c$ axis
(2.08 \AA, average of three) becomes longer than the one parallel to the $c$ axis (1.96 \AA).
The calculated spin densities are shown in Fig.~\ref{f:CN0spinden}.
For the HSE result, the spin density shows clear directional preference along
one of the C--Ga bonds, suggesting directional hole localization due to the asymmetric
relaxation. On the other hand, for the PBE result, the spin density is more isotropic
and directional localization is absent. We also checked the magnetic configuration with
the hole localization direction along the $c$ axis and found that this configuration has
higher energy than the one shown in Fig.~\ref{f:CN0spinden}.
In other structures, i.e.\ $\mathrm{C_{Ga}}$ and $\mathrm{C_I}$ as well as all the complex
structures shown in the following subsection, such a clear (Jahn-Teller) distortion is
not observed. Therefore this effect is specific for the $\mathrm{C_N}$ case.

\begin{figure*}
\includegraphics[width=2.8cm]{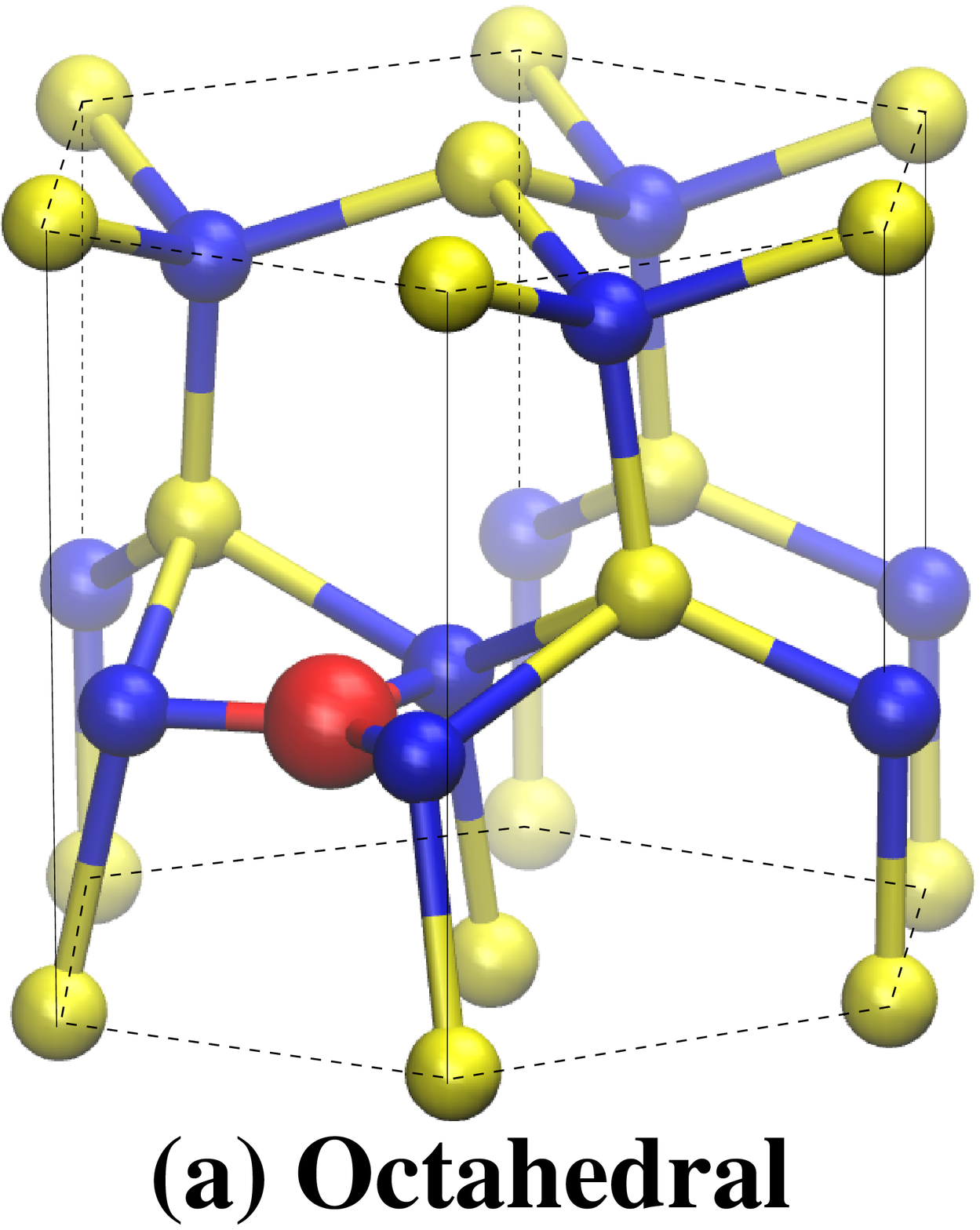}%
\includegraphics[width=3.2cm]{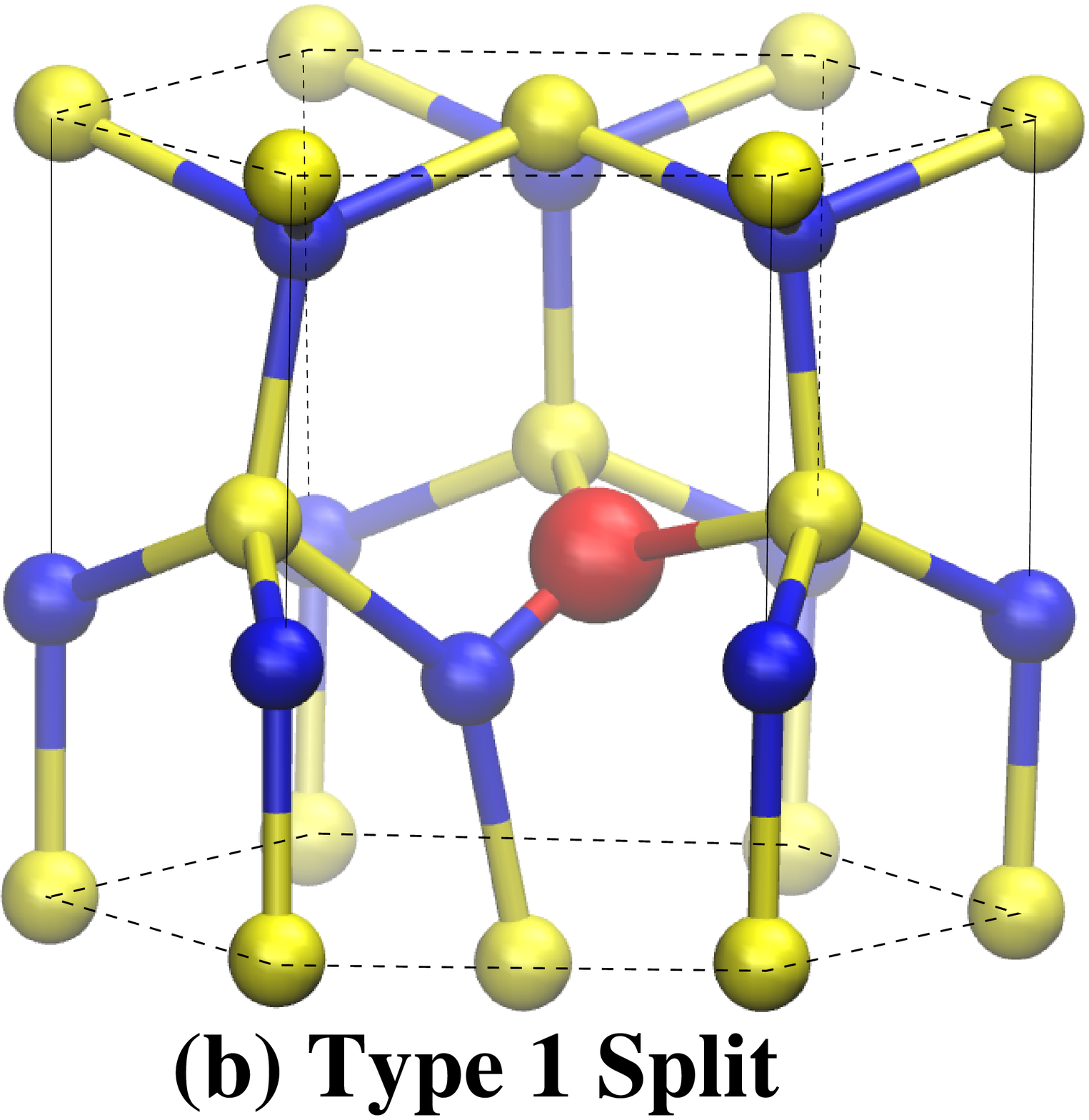}
\includegraphics[width=3.2cm]{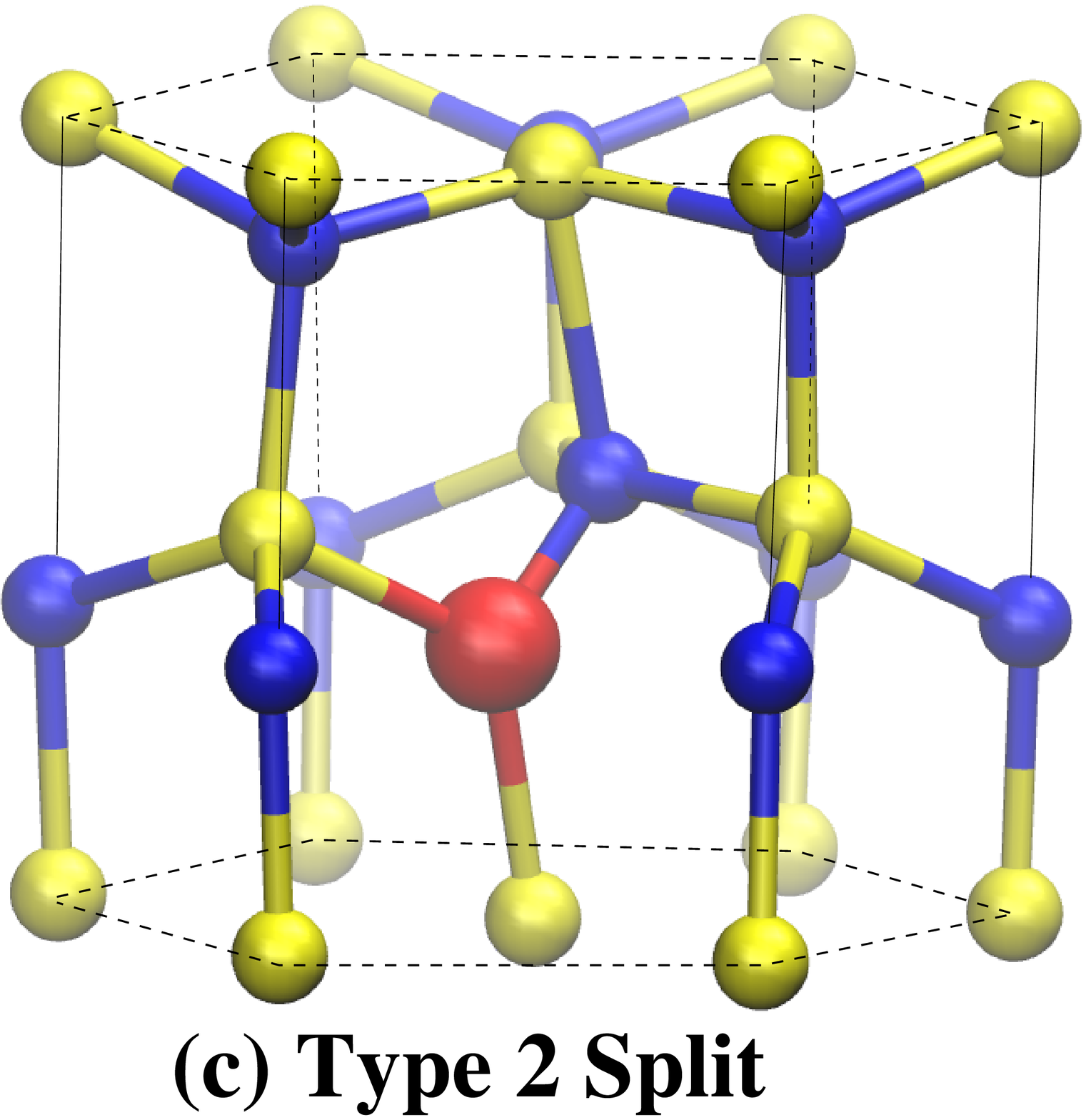}
\includegraphics[width=3.2cm]{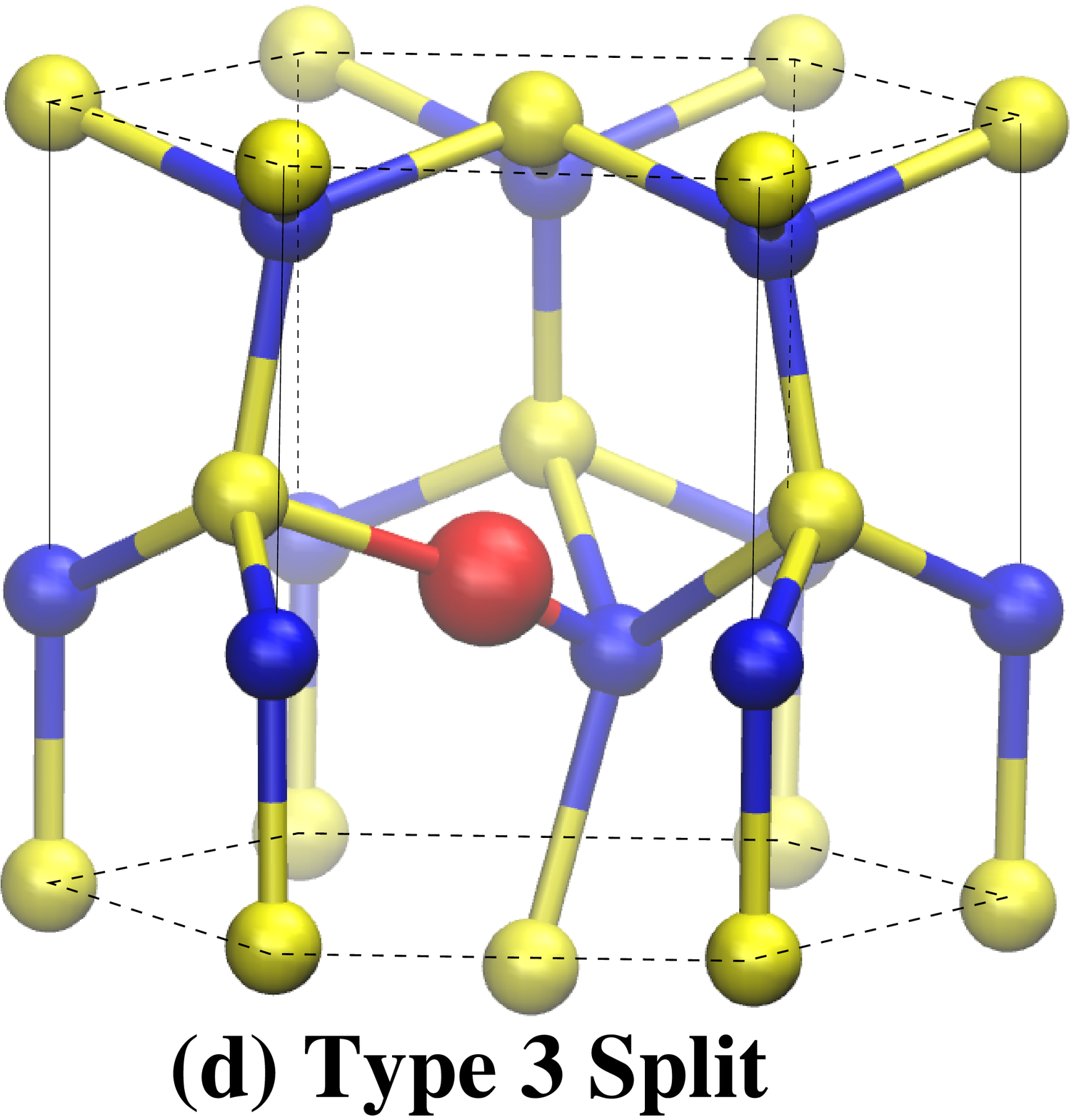}
\includegraphics[width=3.2cm]{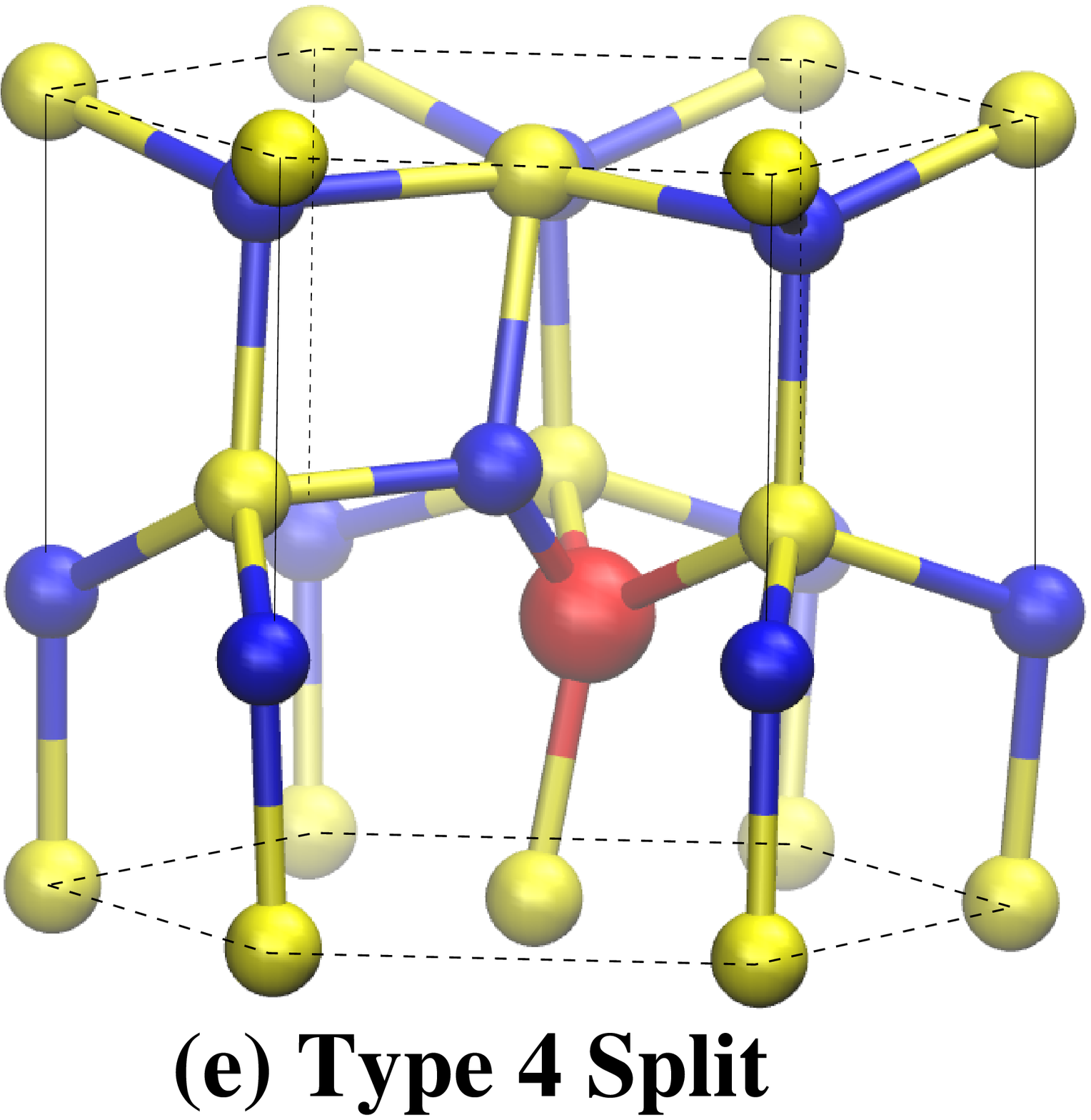}
\caption{\label{f:geoCi}Ball and stick representations of the positions of $\mathrm{C_I}$ in GaN.
C atom is denoted by red sphere, while Ga and N are denoted by yellow and blue spheres, respectively.
Relaxed configurations of (a) octahedral interstitial position with $4+$ charge state, (b) type 1
split interstitial position with $1+$ charge state, (c) type 2 split interstitial position
with $1-$ charge state, (d) type 3 split interstitial position with $2+$ charge state, (e)
type 4 split interstitial positions with $2-$ charge state.}
\end{figure*}

In the case of $\mathrm{C_I}$, there are plenty of possibilities for
the position of C atom as an interstitial in GaN. In this work, we
took the comprehensive study by Wright~\cite{wright02} as our
starting point. We considered an octahedral interstitial position
(denoted as channel configuration in Ref.~[\onlinecite{wright02}]),
tetrahedral interstitial position, split interstitial position and
bond center position as initial configurations of C and then a full
structural optimization was performed for each geometry. After fully
relaxing all the configurations we found that the tetrahedral
position never becomes the most stable and the bond center position
either takes higher formation energy in some charge states or
relaxes into split interstitial positions in other charge states.
Consequently, we focus on the octahedral and split interstitial
positions. In the split interstitial configurations, a tilted C--N
dimer replaces a N atom and, depending on the direction of the
dimer, four different types were considered~\cite{wright02}. In type 1
and 4 split interstitial configurations, C takes higher and
lower positions than N, respectively, and has two bonds with Ga
atoms. In type 2 and 3, C takes lower and higher positions
than N, respectively, and has one bond with a Ga atom. The structures
of these five configurations (octahedral and four split
interstitials) are shown in Fig.~\ref{f:geoCi}. The calculated
formation energies for these $\mathrm{C_I}$ configurations are also
reported in Fig.~\ref{f:EfC} together with substitutional cases.

%
\begin{table*}
\caption{Comparison of the calculated structural and electronic properties of $\mathrm{C_I}$ in the different charge states $q$ between the previous result in LDA~\cite{wright02} and this work in HSE. The preferred forms with lowest energies in each charge state, the C--N bond lengths ($l_{\mathrm{C-N}}$ in \AA) and the formation energies ($E_f$ in eV) are presented.\label{t:CI}}
\begin{ruledtabular}
\begin{tabular}{ccccccc}
 & \multicolumn{3}{c}{LDA~\footnotemark[1]} & \multicolumn{3}{c}{HSE (this work)} \\ \hline
 $q$  & form & $l_{\mathrm{C-N}}$ & $E_f$ & form & $l_{\mathrm{C-N}}$ & $E_f$ \\
 $4+$ & octahedral & $1.40$ & $1.76+4E_F$ & octahedral & $1.37$ & $0.29+4E_F$ \\
 $3+$ & octahedral & --     & $3.60+3E_F$ & octahedral & $1.45$ & $2.16+3E_F$ \\
 $2+$ & split 1    & --     & $3.49+2E_F$ & split 3 & $1.16$ & $1.92+2E_F$ \\
 $1+$ & split 1    & --     & $4.67+E_F$ & split 1 & $1.23$ & $3.54+E_F$ \\
 $0$  & split 1    & $1.33$ & $6.55$ & split 2 & $1.31$ & $6.12$ \\
 $1-$ & split 3    & --     & $8.92-E_F$ & split 2 & $1.39$ & $9.31-E_F$ \\
 $2-$ & split 3    & --     & $11.26-2E_F$ & split 4 & $1.47$ & $13.69-2E_F$ \\
\end{tabular}
\end{ruledtabular}
\footnotetext[1]{Ref.~\onlinecite{wright02}.}
\end{table*}
%

We can notice first that, the octahedral interstitial configuration,
shown in Fig.~\ref{f:geoCi}(a), is the most stable in the $4+$
charge state~\cite{4plus},
which, in turn, is more stable than substitutional
cases, both in N-rich and Ga-rich conditions when the Fermi energy
is located close to the VBM. In this configuration a C atom is
surrounded by three N atoms, which are attracted by the positively
charged C atom.
When considering the $3+$ charge state, the octahedral interstitial
configuration is also more stable than the split interstitial
configurations, but this charge state never becomes favorable within
the band gap. In the case of other charge states, split interstitial
configurations are favorable.

The four variants of split interstitial configurations have very
similar formation energies in each charge state. In the $2+$ charge
state, type 3 split interstitial configuration, in
Fig.~\ref{f:geoCi}(d), is the most favorable.
In the $1+$ charge state, type 1 split interstitial
configuration [Fig.~\ref{f:geoCi} (b)] is the most stable.
In the case of 0 (neutral)
charge state, type 2 split interstitial configuration is the most
stable, but type 1 split interstitial configurations have almost
identical formation energies within a 10\,meV range.
%
In the case of $1-$ charge state, type 2 split
interstitial configuration is the most stable.
Type 4 split interstitial configuration becomes
favorable in the $2-$ charge state, but this state is never stable
within the band gap.

\begin{figure}
\includegraphics[width=\columnwidth]{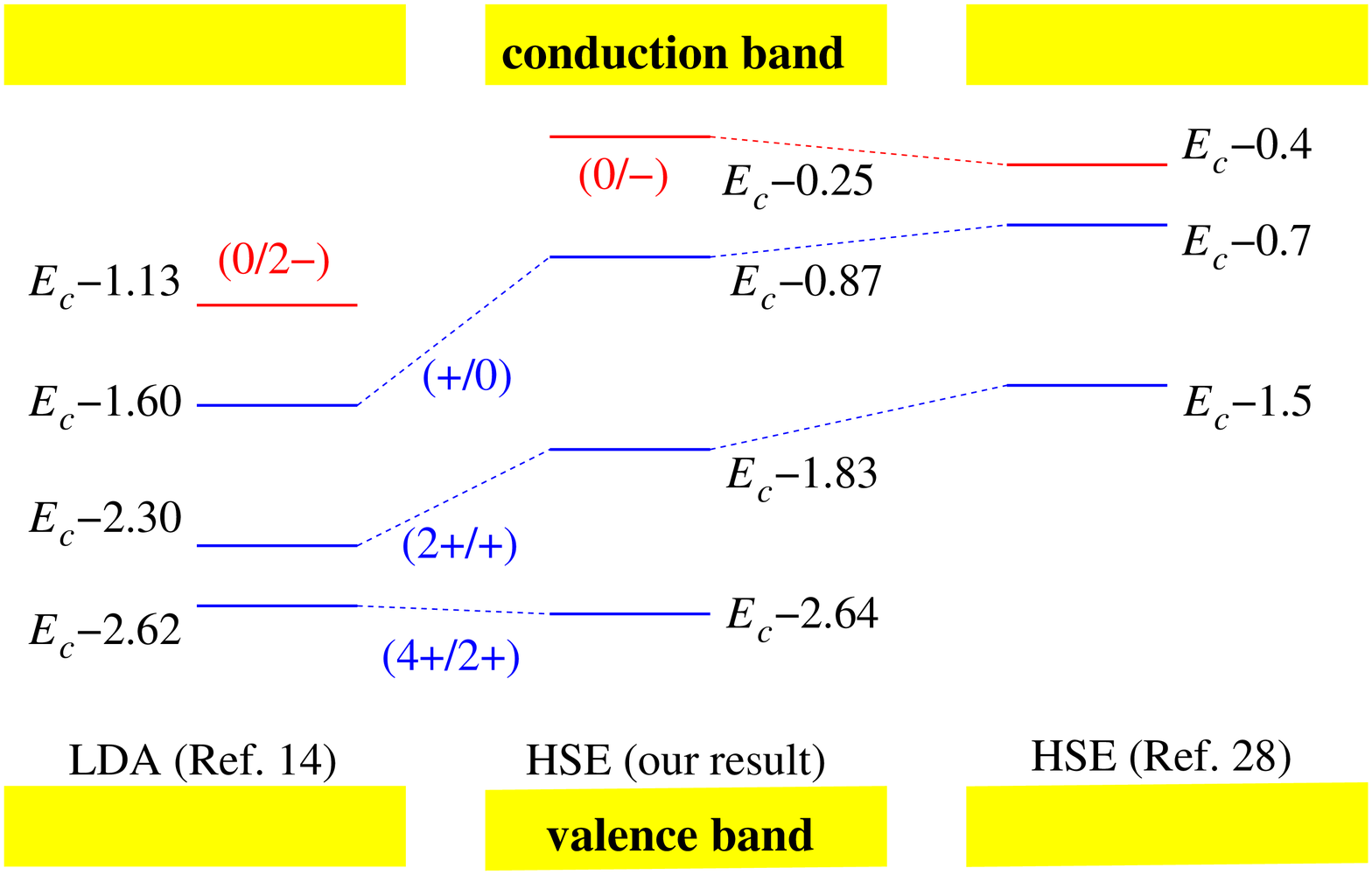}%
\caption{\label{f:Cilevel} Thermodynamic transition levels for $\mathrm{C_I}$ obtained by LDA~\cite{wright02} (in left) and by HSE (our results in center and from Ref.~\onlinecite{lyons14} in right).
The positions of the levels are measured from the conduction band minimum ($E_c$). Four transition levels appear in LDA and our HSE results, whereas the (4+/2+) transition level is absent in the results from Ref.~\onlinecite{lyons14}. Three of four levels in LDA and our HSE results
correspond to the same transitions, (4+/2+), (2+/+) and (+/0). The level
closest to the $E_c$ is (0/$2-$) in LDA result, whereas it is (0/$-$) in both HSE results.
Note that no band edge alignment is considered.}
\end{figure}

To conclude this section, Table~\ref{t:CI} and Fig.~\ref{f:Cilevel} provide a summary of
our results for $\mathrm{C_I}$ and we compare them with previously
obtained LDA results~\cite{wright02} and recently obtained HSE results~\cite{lyons14}.
In Fig.~\ref{f:Cilevel}, it can be seen that, in
the case of LDA, the (0/2$-$) transition occurs at $E_c -1.13$\,eV,
but in both HSE results the $2-$ charge state is never stable within band gap and the
(0/$-$) transition level appears at $E_c - 0.25$\,eV (our result), instead.
Furthermore, the transition levels for (+/0) and (2+/+) obtained
with HSE are shifted closer to CBM, whereas (4+/2+) level shows
small shift closer to VBM. In the previous HSE result~\cite{lyons14}, 4+ charge state is not reported.
Therefore the (4+/2+) transition level is absent. The positions of three other levels between our results and the results in Ref.~\onlinecite{lyons14} are different up to $\sim$0.35\,eV. The reason is not clear, but we may attribute it to the use of different pseudopotentials and cutoff energy.
For the comparison of LDA and HSE results in Fig.~\ref{f:Cilevel}, band edge alignment procedure~\cite{alkauskas08,alkauskas11prb,miceli15} was not performed due to the unavailability of the details of band gap correction procedure in Ref.~\onlinecite{wright02}. It is possible that there exists substantial amount of valence band off-set between LDA and HSE results.


\subsection{Carbon complexes\label{ss:complex}}

\begin{figure}
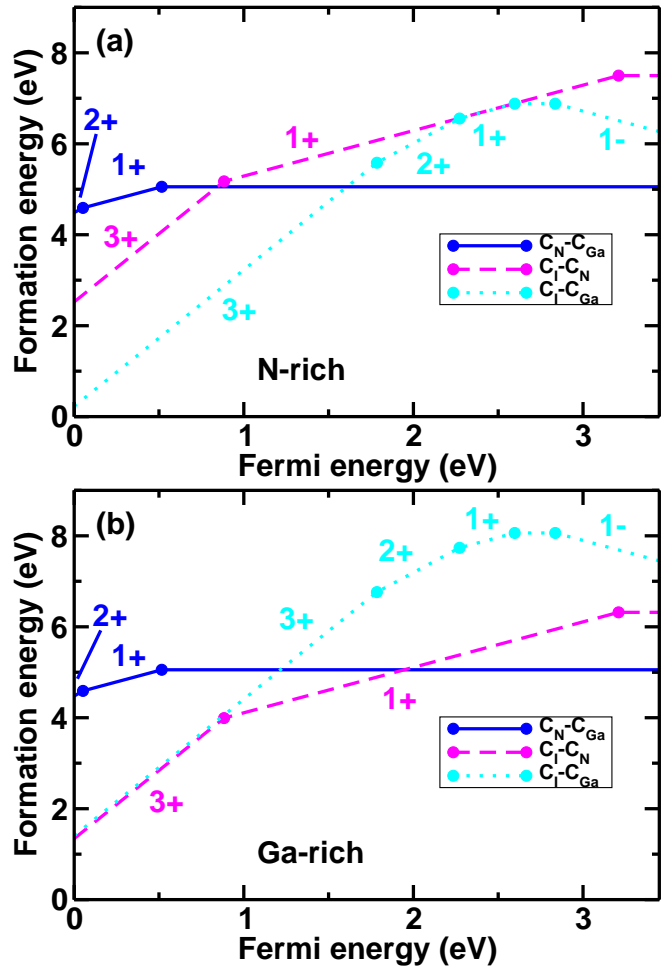

\includegraphics[width=\columnwidth]{fCc-Nrich.eps} \\
\includegraphics[width=\columnwidth]{fCc-Garich.eps}%
\caption{\label{f:EfCc}Formation energies as a function of Fermi energy for $\mathrm{C_N-C_{Ga}}$ (blue solid line), $\mathrm{C_I-C_N}$ (magenta dashed line) and $\mathrm{C_I-C_{Ga}}$ (cyan dotted line) in (a) N-rich conditions and (b) Ga-rich conditions.
}
\end{figure}

\begin{figure}
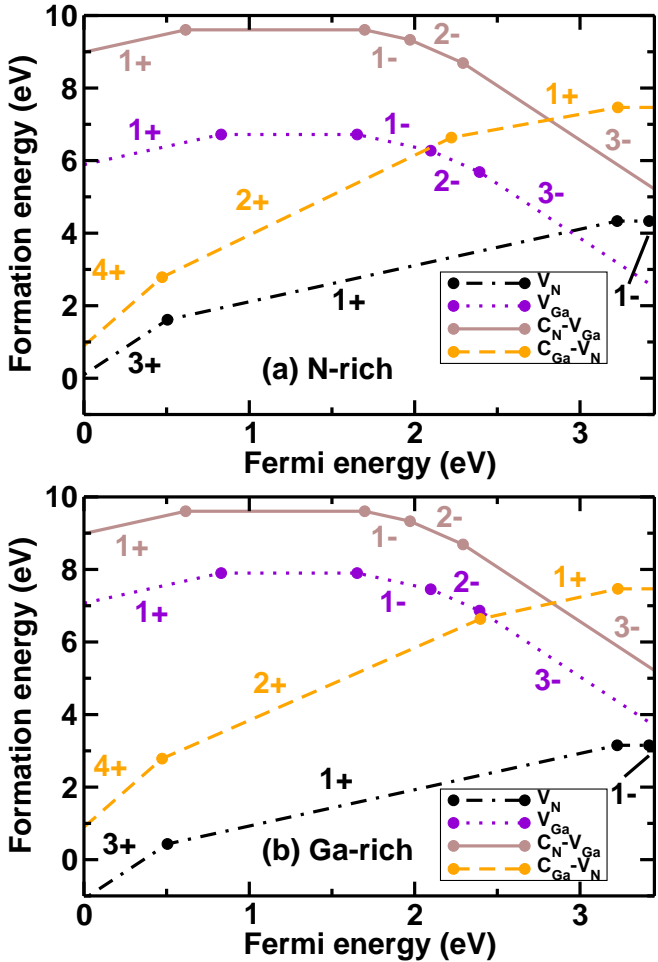

\includegraphics[width=\columnwidth]{fVc-Nrich.eps} \\
\includegraphics[width=\columnwidth]{fVc-Garich.eps}%
\caption{\label{f:EfVc}Formation energies as a function of Fermi energy for $\mathrm{C_N-V_{Ga}}$ (solid line in brown) and $\mathrm{C_{Ga}-V_N}$ (dashed line in orange)
in (a) N-rich conditions and (b) Ga-rich conditions.
Formation energies for $\mathrm{V_N}$ (dashed-dotted line in black) and
$\mathrm{V_{Ga}}$ (dotted line in violet) are also plotted for comparison.
}
\end{figure}

For the carbon complexes we consider combinations of single impurity carbon, i.e.
$\mathrm{C_N-C_{Ga}}$, $\mathrm{C_I-C_N}$ and $\mathrm{C_I-C_{Ga}}$.
Formation energies of these complexes as a function of Fermi level are shown in
Fig.~\ref{f:EfCc}.
Additionally, complexes of substitutional C with neighboring vacancies,
i.e. $\mathrm{C_N-V_{Ga}}$ and $\mathrm{C_{Ga}-V_N}$ are also considered.
Formation energies of these complexes are shown in Fig.~\ref{f:EfVc}.


\subsubsection{Binding energy\label{sss:EB}}

\begin{figure}
\includegraphics[width=\columnwidth]{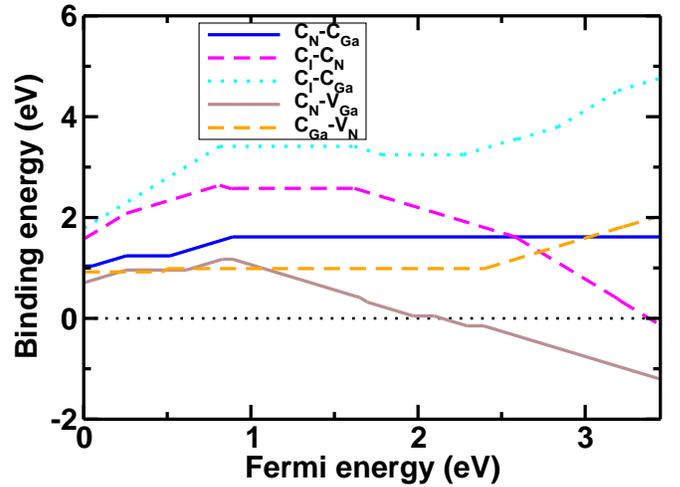}%
\caption{\label{f:EB}Binding energies as a function of Fermi energy for $\mathrm{C_N-C_{Ga}}$ (solid line in blue), $\mathrm{C_I-C_N}$ (dashed line in magenta), $\mathrm{C_I-C_{Ga}}$ (dotted line in cyan), $\mathrm{C_N-V_{Ga}}$ (solid line in brown) and $\mathrm{C_{Ga}-V_N}$ 
(dashed line in orange).}
\end{figure}

For these carbon complexes we calculate binding energies in addition to formation energies.
The binding energy ($E_{\mathrm{B}}$) for the complex denoted by $A-B$ is defined as
\begin{eqnarray}
E_{\mathrm{B}}^{q_A+q_B-q_{A-B}} (A-B, E_F) &=& E_f^{q_A} (A, E_F) + E_f^{q_B} (B, E_F) \nonumber \\
&-& E_f^{q_{A-B}} (A-B, E_F),
\end{eqnarray}
where the formation energies are chosen as the lowest energy
configuration of each defect state at a particular Fermi energy.
With this definition, the complex is stable (unstable), when
$E_{\mathrm{B}}$ takes positive (negative) value. The calculated
binding energy for $\mathrm{C_N-C_{Ga}}$, $\mathrm{C_I-C_N}$,
$\mathrm{C_I-C_{Ga}}$, $\mathrm{C_N-V_{Ga}}$ and $\mathrm{C_{Ga}-V_N}$ are shown in Fig.~\ref{f:EB}.

\subsubsection{$\mathrm{C_N-C_{Ga}}$}

\begin{figure}
\includegraphics[width=4cm]{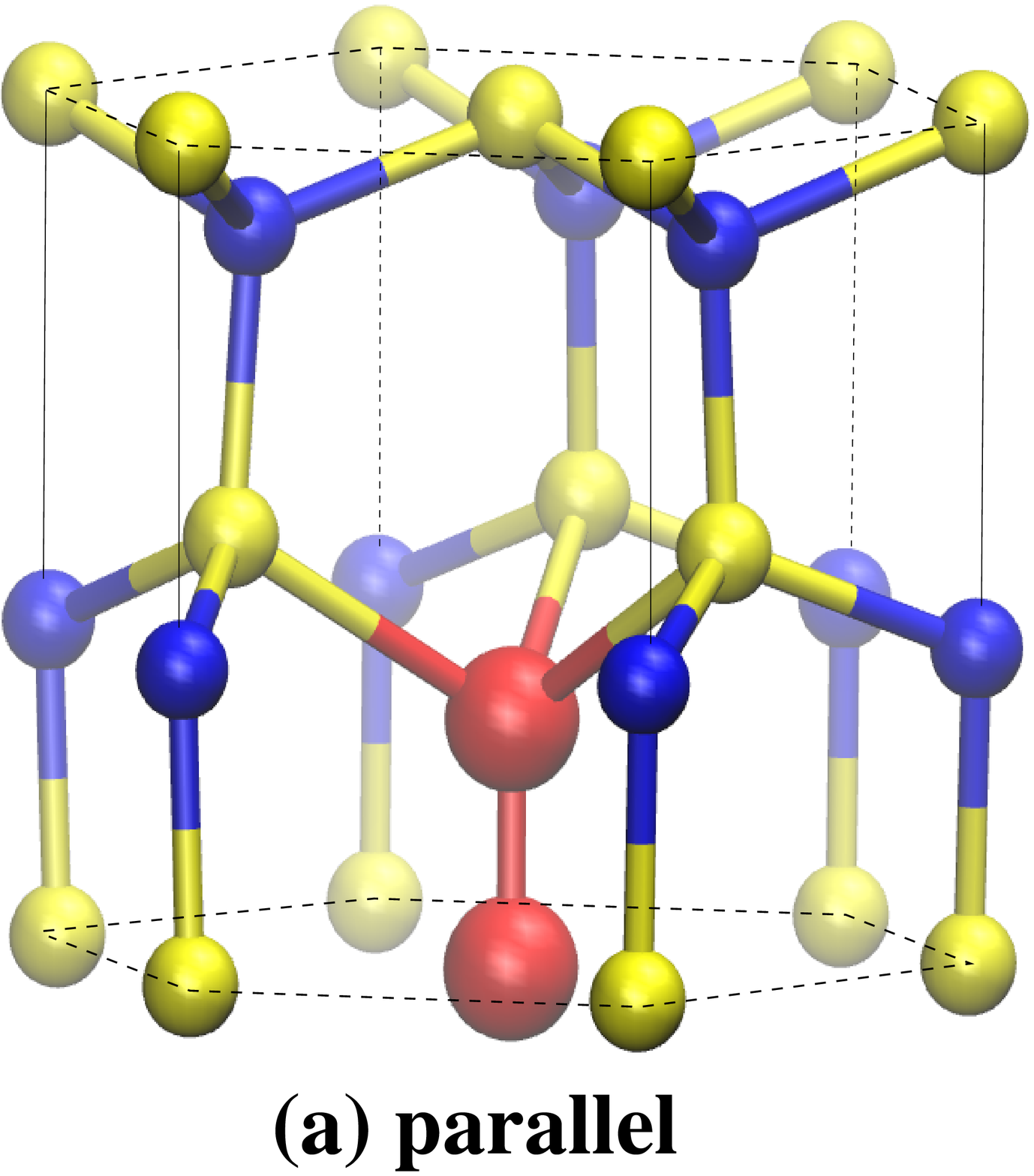}%
\includegraphics[width=4cm]{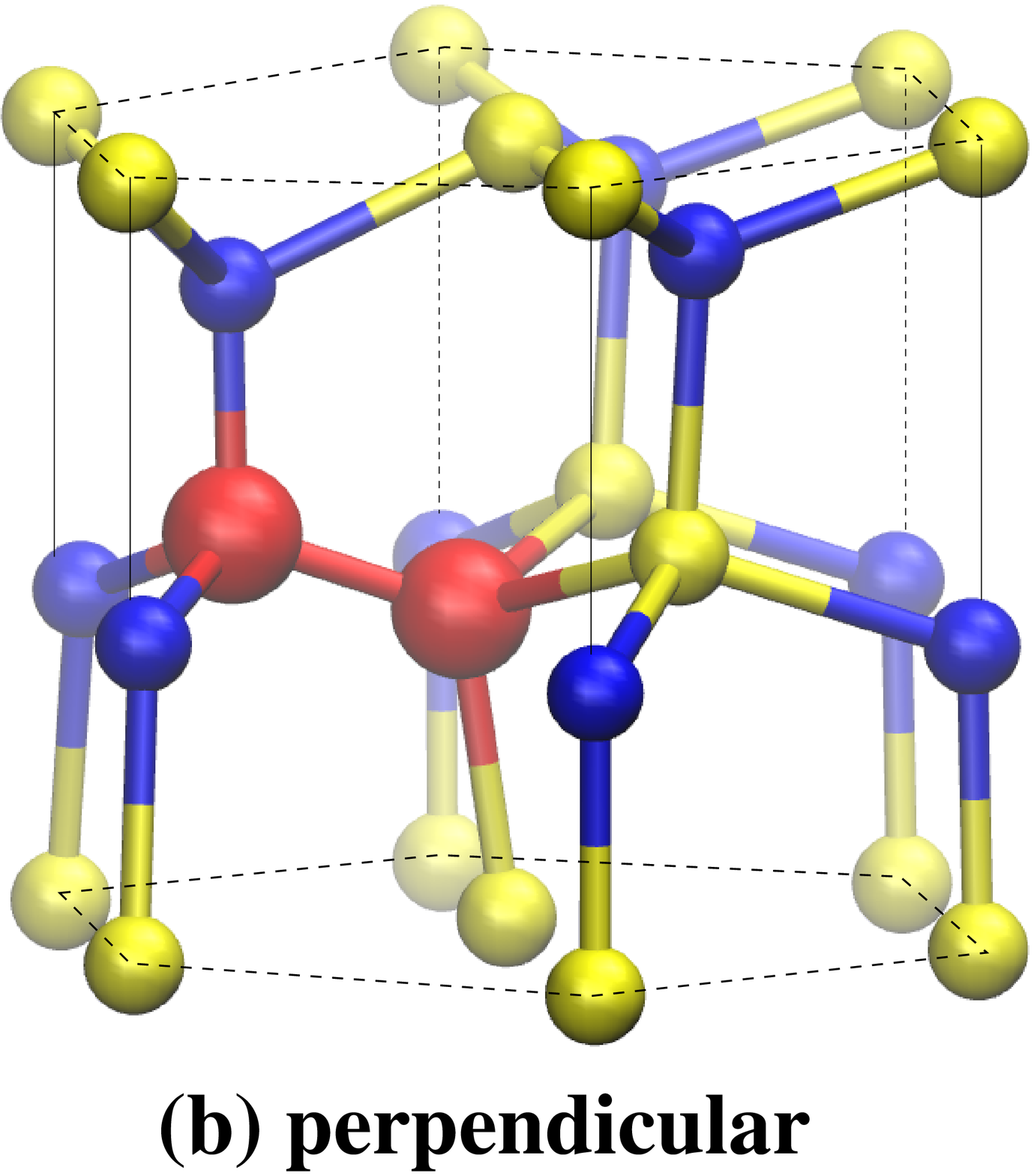}%
\caption{\label{f:geoCNCGa}Ball and stick representations of $\mathrm{C_N-C_{Ga}}$ complex.
(a) $\mathrm{C_N}$ and $\mathrm{C_{Ga}}$ are located parallel to the $c$-axis. (b) $\mathrm{C_N}$
and $\mathrm{C_{Ga}}$ are located perpendicular to the $c$-axis.}
\end{figure}

For this kind of complex two different configurations are possible
and their relaxed structures are shown in Fig.\,\ref{f:geoCNCGa}.
In one configuration, two C atoms are located along the $c$-axis and
we will refer to it as the parallel configuration, as shown in
Fig.\,\ref{f:geoCNCGa}(a). In the other configuration, two C atoms
are located nearly perpendicular to the $c$-axis and this is
referred to as the perpendicular configuration shown in
Fig.\,\ref{f:geoCNCGa}(b). Formation energies are calculated for
both configurations and we find that the difference between them is
very small, less than 0.1\,eV, although the perpendicular
configuration has lower formation energies than the parallel
configuration.
The formation energy of the perpendicular configuration
$\mathrm{C_N-C_{Ga}}$ complex is plotted in Fig.\,\ref{f:EfCc} with
blue solid lines. Three charge states are favorable within the band gap.
Up to 0.05\,eV, the $2+$ charge state is the most stable.
Then for values of the Fermi energy up to 0.52\,eV the 1+
charge state is the most stable. The neutral charge state instead,
is the most stable when the Fermi energy is above 0.52\,eV. Unlike
previous LDA result~\cite{wright02}, negatively charged states are
not present in the band gap.
The C--C bond lengths and formation energies in each charge state are summarized
in Table~\ref{t:CNCGa} together with previous LDA results~\cite{wright02} for
comparison.
Moreover, with a binding energy in excess of 1\,eV (see
Fig.~\ref{f:EB}), $\mathrm{C_N}$ and $\mathrm{C_{Ga}}$ form stable
complexes in GaN. In the upper half of the fundamental band gap, the
neutral charge state of this complex is the most favorable among all
the C-complexes considered here, both in N-rich and Ga-rich limits.
It should be noted that both $\mathrm{C_{Ga}}$ and $\mathrm{C_N}$
are positively charged near the VBM and are expected to repel each other.
This may hinder the formation of the 2+ charge state of $\mathrm{C_N-C_{Ga}}$ complex.

%

%
\begin{table*}
\caption{Comparison of the calculated structural and electronic properties of $\mathrm{C_N-C_{Ga}}$ in the different charge states $q$ between the previous result in LDA~\cite{wright02} and this work in HSE. The preferred forms, parallel ($\parallel$) or perpendicular ($\perp$) configurations, with lowest energies in each charge state, the C--C bond lengths ($l_{\mathrm{C-C}}$ in \AA) and the formation energies ($E_f$ in eV) are presented.\label{t:CNCGa}}
\begin{ruledtabular}
\begin{tabular}{ccccccc}
 & \multicolumn{3}{c}{LDA~\footnotemark[1]} & \multicolumn{3}{c}{HSE (this work)} \\ \hline
 $q$  & form & $l_{\mathrm{C-C}}$ & $E_f$ & form & $l_{\mathrm{C-C}}$ & $E_f$ \\
 $2+$ & --          & --    & --         & $\perp$ & $1.53$ & $4.49+2E_F$ \\
 $1+$ & $\parallel$ & --    & $4.74+E_F$ & $\perp$ & $1.55$ & $4.54+E_F$ \\
 $0$  & $\perp$     & $1.5$ & $4.81$     & $\perp$ & $1.56$ & $5.05$ \\
 $1-$ & $\perp$     & --    & $8.19-E_F$ & $\perp$ & $1.57$ & $9.25-E_F$ \\
\end{tabular}
\end{ruledtabular}
\footnotetext[1]{Ref.~\onlinecite{wright02}.}
\end{table*}
%

\subsubsection{$\mathrm{C_I-C_N}$}

\begin{figure}
\includegraphics[width=2.5cm]{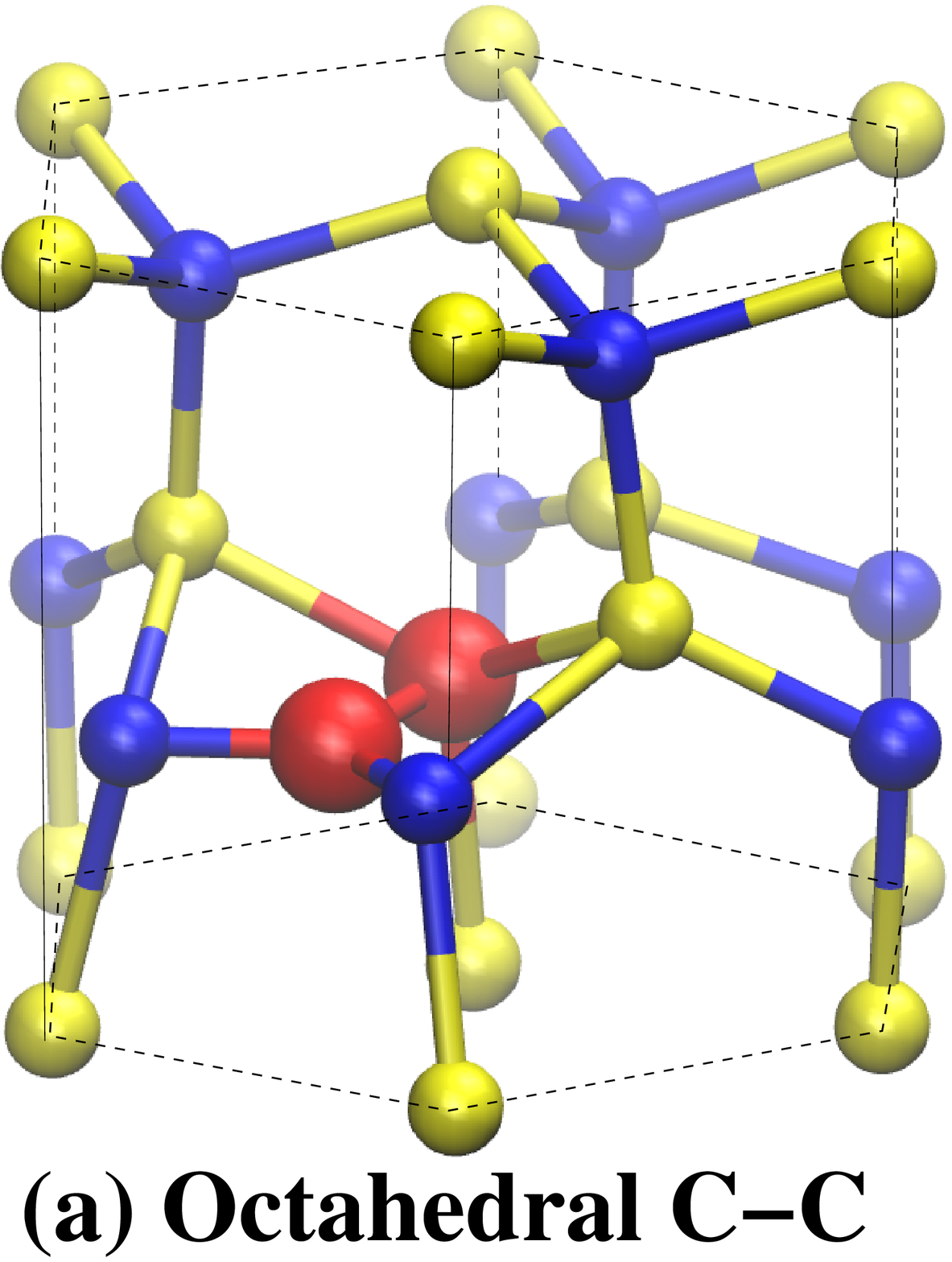}%
\includegraphics[width=2.75cm]{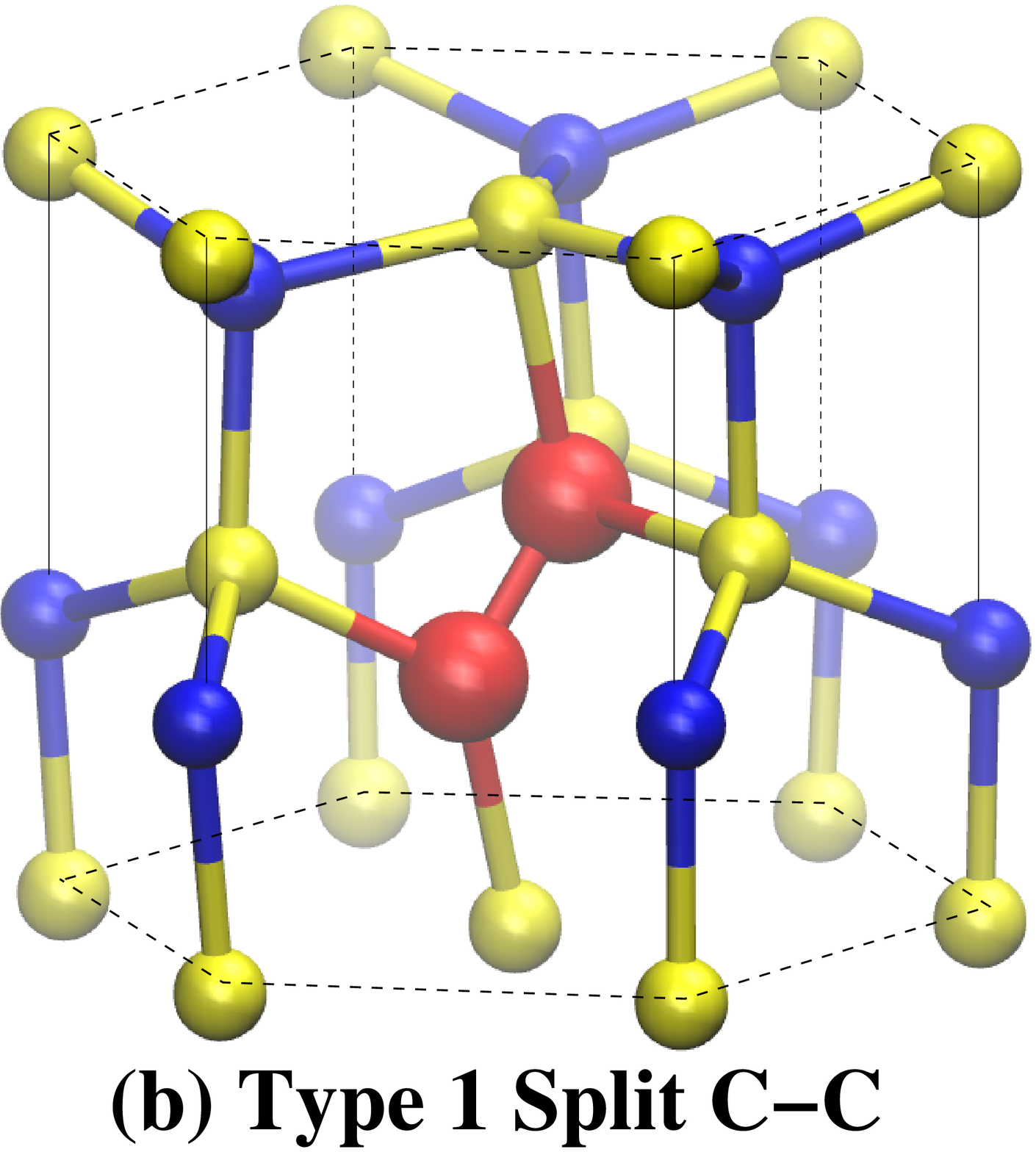}%
\includegraphics[width=2.75cm]{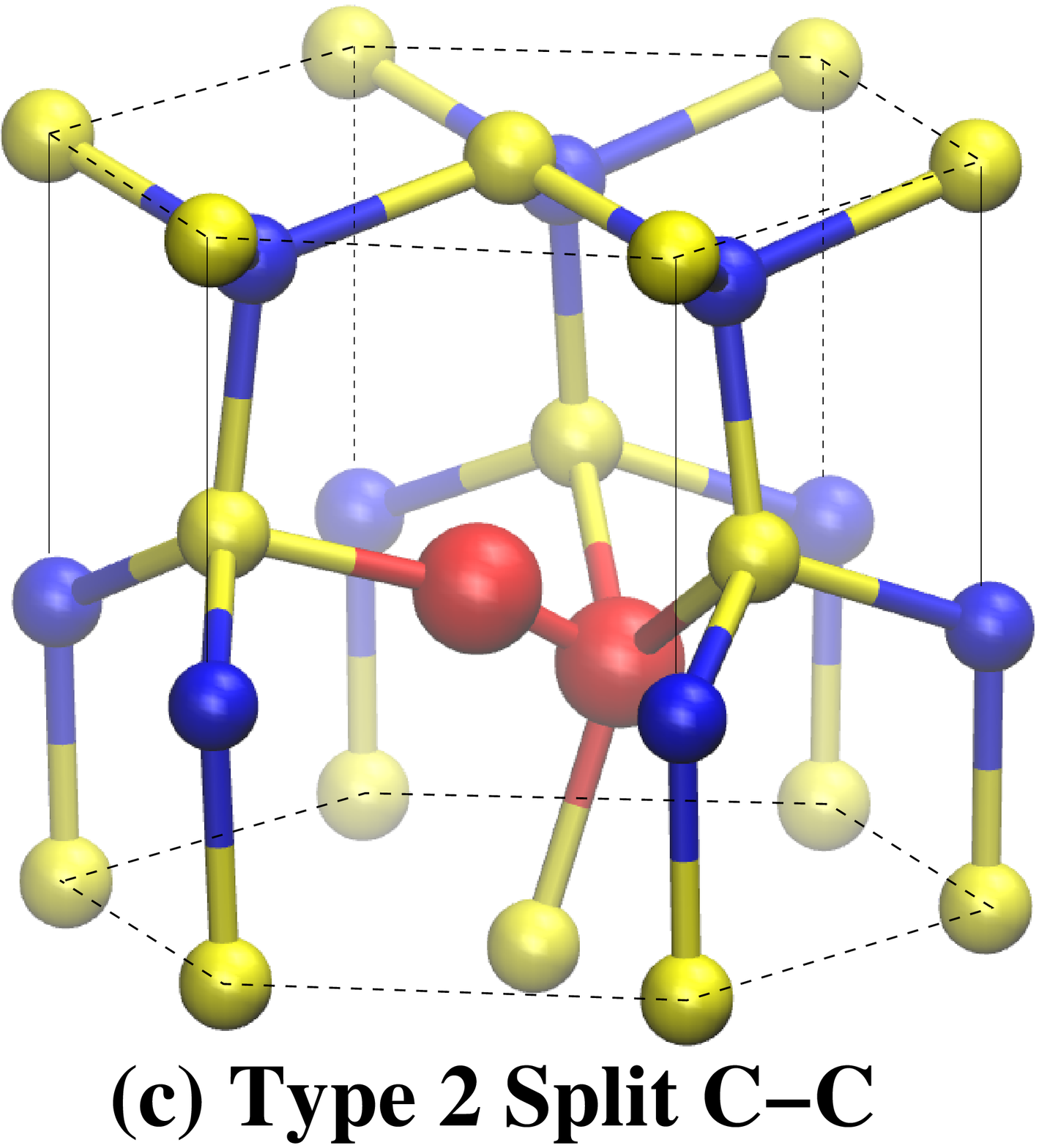}%
\caption{\label{f:geoCICN}Ball and stick representations of $\mathrm{C_I-C_N}$ complex.
(a) Octahedral C--C configuration: One of the C atoms is located at the octahedral interstitial
position and the other is $\mathrm{C_N}$.
(b) Type 1 split C--C configuration: N atom of C--N dimer in Type 1 (and Type 2) split interstitial
of $\mathrm{C_I}$ is replaced by C.
(c) Type 2 split C--C configuration: N atom of C--N dimer in Type 3 (and Type 4) split
interstitial of $\mathrm{C_I}$ is replaced by C.
}
\end{figure}

Three different configurations are considered for the
$\mathrm{C_I-C_N}$ complex and they are shown in
Fig.~\ref{f:geoCICN}. The first configuration is the combination of
octahedral configuration from $\mathrm{C_I}$ and $\mathrm{C_N}$
as shown in Fig.~\ref{f:geoCICN}(a). The second configuration is
composed of a Type 1 (Type 2) split interstitial $\mathrm{C_I}$ and
$\mathrm{C_N}$, where the high-positioned C atom has two bonds with
Ga atoms. This is shown in Fig.~\ref{f:geoCICN}(b). Finally, the
third one is a combination of Type 3 (and Type 4) split interstitial
$\mathrm{C_I}$ and $\mathrm{C_N}$, where the high-positioned C atom
has one bond with a Ga atom, as indicated in
Fig.~\ref{f:geoCICN}(c).

The formation energy values obtained for the $\mathrm{C_I-C_N}$
complex are plotted (magenta dashed lines) in Fig.~\ref{f:EfCc}. The 3+, 1+
and neutral (0) charge are characterized by states positioned in the
band gap. The 3+ charge state is favorable for Fermi
energies up to 0.88\,eV, whereas the 1+ charge state is stable
between 0.88\,eV and 3.21\,eV. Consequently, this complex, in 1+
state, mostly acts as a deep donor.

Additionally, in the Ga-rich limit, the 1+ charge state becomes the
most favorable form among all the C-complex considered here, up to
mid-gap. On the other hand, in the N-rich limit, this complex is
never favorable. Fig.~\ref{f:EB} shows that binding energy for this
complex is decreasing when the Fermi level is approaching the CBM
and eventually becomes negative at around 3.38\,eV.

In the 3+ charge state, this complex assumes an octahedral C--C
configuration as indicated in Fig.~\ref{f:geoCICN}(a)
. Both the 1+ and 0 (neutral) charge
states, are found to be in the Type\,2 C--C interstitial
configuration, as shown in Fig.~\ref{f:geoCICN}(c)
.
The 1+ charge state of $\mathrm{C_I-C_N}$ is also studied in Ref.~\onlinecite{lyons14}.
The reported values of 2.62 eV binding energy and 1.23\,\AA\ of C--C bond length are in
good agreement with our results.

The C--C bond lengths and formation energies in each charge state are summarized
in Table~\ref{t:CICN} together with previous LDA results~\cite{wright02} for
comparison.

%

%
\begin{table*}
\caption{Comparison of the calculated structural and electronic properties of $\mathrm{C_I-C_N}$ in the different charge states $q$ between the previous result in LDA~\cite{wright02} and this work in HSE. The preferred forms with lowest energies in each charge state, the C--C bond lengths ($l_{\mathrm{C-C}}$ in \AA) and the formation energies ($E_f$ in eV) in Ga-rich conditions are presented.\label{t:CICN}}
\begin{ruledtabular}
\begin{tabular}{ccccccc}
 & \multicolumn{3}{c}{LDA~\footnotemark[1]} & \multicolumn{3}{c}{HSE (this work)} \\ \hline
 $q$  & form & $l_{\mathrm{C-C}}$ & $E_f$ & form & $l_{\mathrm{C-C}}$ & $E_f$ \\
 $3+$ & octahedral & $1.41$ & $2.65+3E_F$ & octahedral & $1.43$ & $1.34+3E_F$ \\
 $1+$ & split 1    & $1.24$ & $3.88+E_F$  & split 2    & $1.29$ & $3.11+E_F$ \\
 $0$  & --         & --     & --          & split 2    & $1.23$ & $6.32$ \\
\end{tabular}
\end{ruledtabular}
\footnotetext[1]{Ref.~\onlinecite{wright02}.}
\end{table*}
%

\subsubsection{$\mathrm{C_I-C_{Ga}}$}

\begin{figure}
\includegraphics[width=2.55cm]{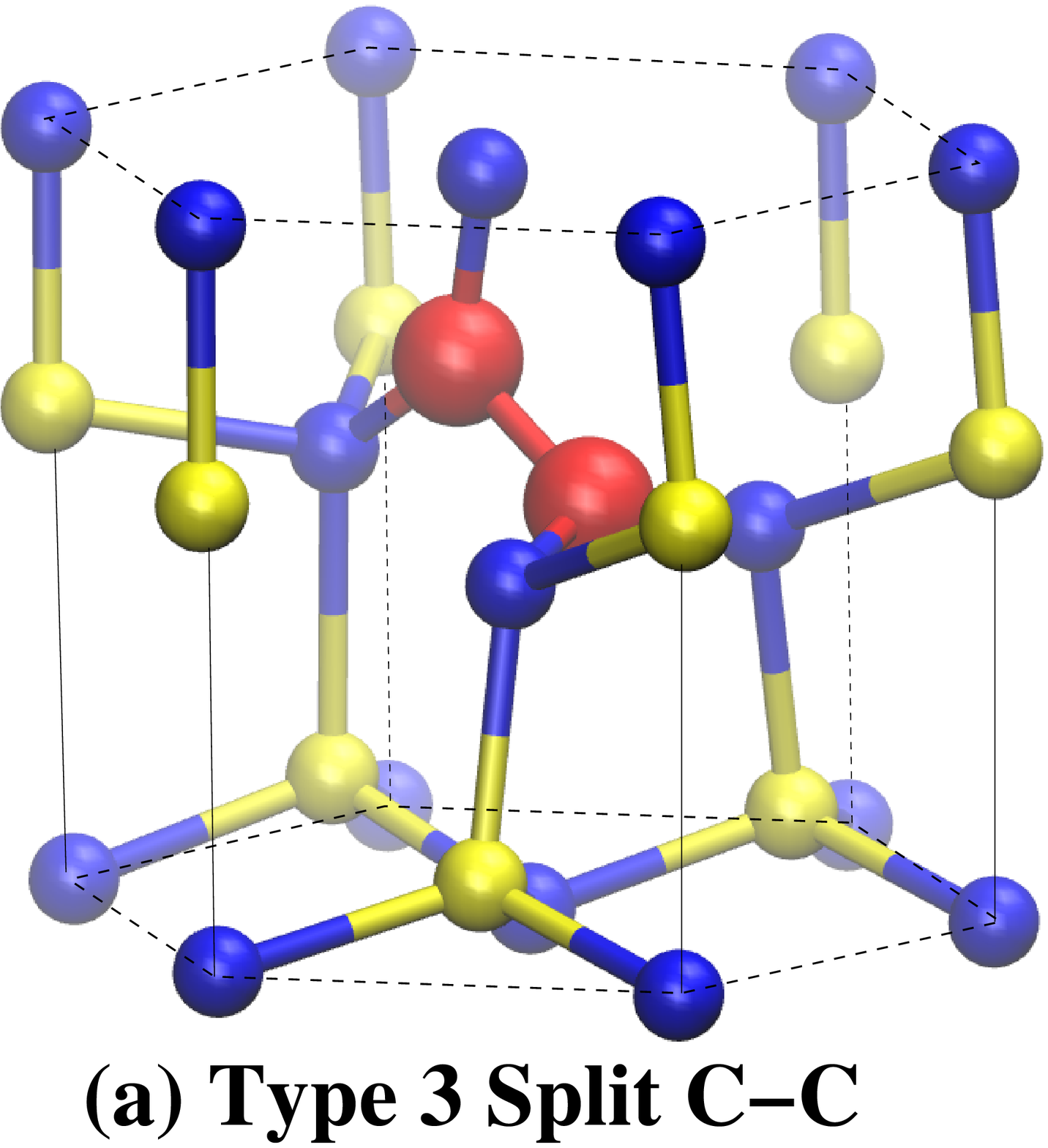}%
\includegraphics[width=2.75cm]{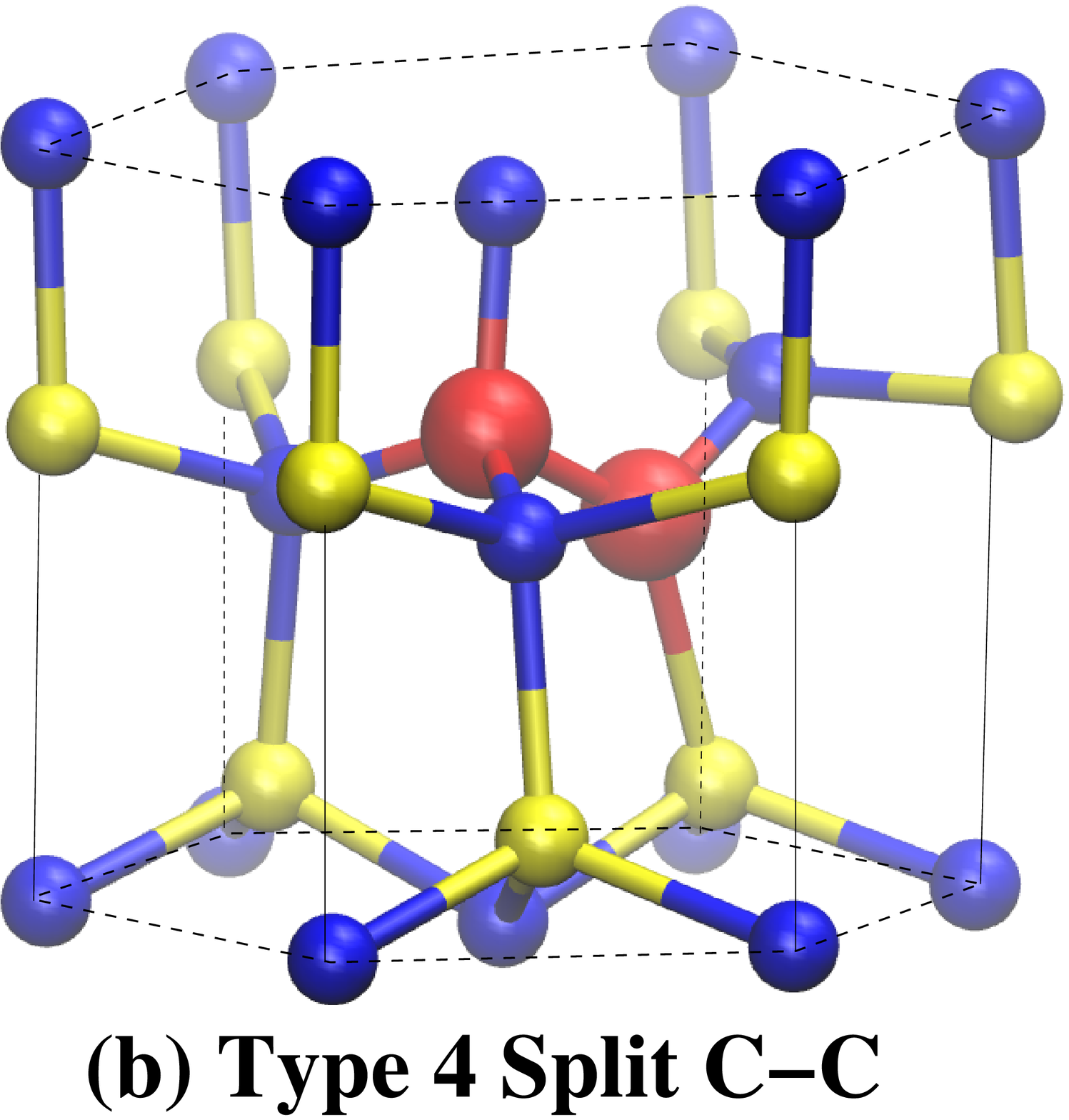}%
\includegraphics[width=2.55cm]{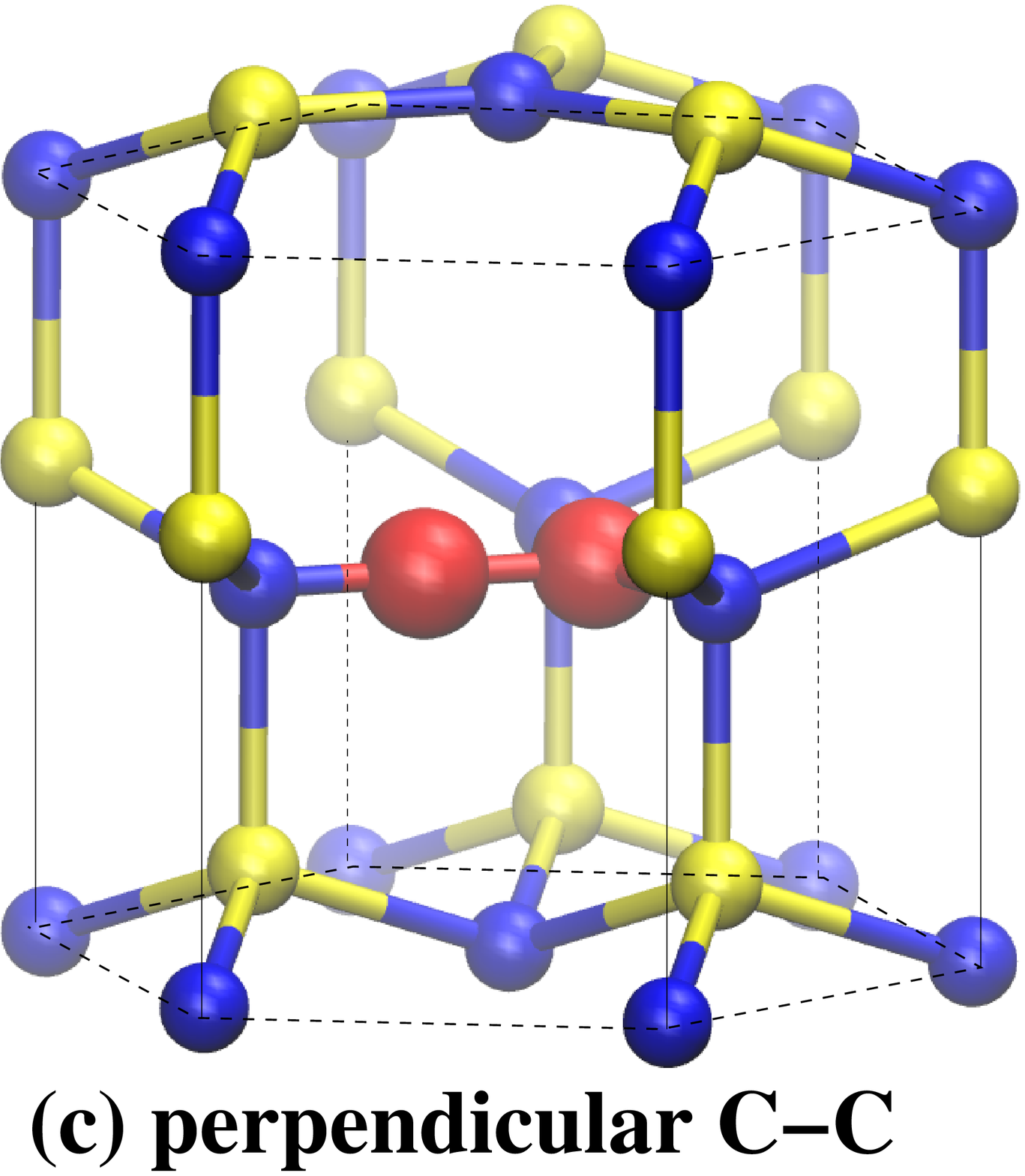}%
\caption{\label{f:geoCICGa}Ball and stick representations of $\mathrm{C_I-C_{Ga}}$ complex.
a tilted C--C dimer replaces a Ga atom.
(a) Type 3 split C--C configuration: both higher and lower C atoms have two
bonds with surrounding N atoms.
(b) Type 4 split C--C configuration: higher C atom has three bonds and the lower
C atom has one bond with surrounding N atoms. (c) The C--C dimer is perpendicular
to the $c$-axis and the N atom above the dimer takes a planar configuration with the
surrounding three Ga atoms.
}
\end{figure}

Two different configurations are obtained after relaxing the
structure of the $\mathrm{C_I-C_{Ga}}$ complex. One configuration is
associated to the 3+, 2+ and 1+ charge states, in which a C--C dimer
replaces a Ga atom and both {high-} and {low-positioned} C atoms
have two bonds with surrounding N atoms. This is a Type\,3 split
C--C configuration as shown in Fig.~\ref{f:geoCICGa}(a). The other
configuration is obtained for the neutral and $1-$ charge states, in
which the high-positioned C atom has three bonds and the
low-positioned one has one bond with surrounding N atoms. This is a
Type 4 split C--C configuration as depicted in
Fig.~\ref{f:geoCICGa}(b). In Ref.~\onlinecite{wright02}, another
configuration, presented in Fig.~\ref{f:geoCICGa}(c), was reported
to be a stable structure, but in our calculation we find that it
never becomes energetically favorable. Formation energy values
obtained for the $\mathrm{C_I-C_{Ga}}$ complex are plotted (dotted
lines) in Fig.~\ref{f:EfCc}. The 3+, 2+, 1+, 0 (neutral) and $1-$
charge states are available for values of the Fermi energy within
the band gap. Moreover, the 3+ charge state is favorable for
energies up to 1.79\,eV.
In N-rich limit, the $\mathrm{C_I-C_{Ga}}$ complex with 3+ charge
state is the most favorable among the C-related complexes considered
here. In Ga-rich limit, the 3+ charge state of this complex together
with same charge state of $\mathrm{C_I-C_N}$ are the most favorable
near the valence band maximum. Furthermore, the
$\mathrm{C_I-C_{Ga}}$ 2+ charge state is available up to 2.27\,eV,
followed by the 1+ charge state that is favorable up to 2.60\,eV.
Above 2.60\,eV, the 0 (neutral) charge state is the most favorable.
Finally, the $-1$ charge state becomes the most favorable above
2.84\,eV. Thus, this complex shows amphoteric behavior, similar to
$\mathrm{C_I}$. However, charge states other than the 3+ have higher
formation energies than $\mathrm{C_N-C_{Ga}}$ and/or
$\mathrm{C_I-C_N}$. Fig.~\ref{f:EB} shows that this complex forms
with binding energy around 2\,eV which subsequently increases above
this value for Fermi energies in the upper half of the band gap.
Note that, despite its high binding energy, the formation of this
complex may be hindered, in particular, in $p$-type GaN (lower half of
the band gap), because both $\mathrm{C_{Ga}}$ and $\mathrm{C_I}$ are positively
charged and are expected to repel each other.
The C--C bond lengths and formation energies in each charge state are summarized
in Table~\ref{t:CICGa} together with previous LDA results~\cite{wright02} for
comparison.

%

%
\begin{table*}
\caption{Comparison of structural and electronic properties of $\mathrm{C_I-C_{Ga}}$ in the different charge states $q$ between the previous result in LDA~\cite{wright02} and this work in HSE. The preferred forms with lowest energies in each charge state, the C--C bond lengths ($l_{\mathrm{C-C}}$ in \AA) and the formation energies ($E_f$ in eV) in Ga-rich conditions are presented.\label{t:CICGa}}
\begin{ruledtabular}
\begin{tabular}{ccccccc}
 & \multicolumn{3}{c}{LDA~\footnotemark[1]} & \multicolumn{3}{c}{HSE (this work)} \\ \hline
 $q$  & form & $l_{\mathrm{C-C}}$ & $E_f$ & form & $l_{\mathrm{C-C}}$ & $E_f$ \\
 $3+$ & split 3 & $1.43$ & $1.94+3E_F$ & split 3 & $1.45$ & $1.41+3E_F$ \\
 $2+$ & split 3 & --     & $3.94+2E_F$ & split 3 & $1.43$ & $3.19+2E_F$ \\
 $1+$ & split 3 & --     & $6.22+E_F$  & split 3 & $1.41$ & $5.47+E_F$ \\
 $0$  & $\perp$ & $1.43$ & $8.48$      & split 4 & $1.51$ & $8.06$ \\
 $1-$ & $\perp$ & --     & $10.13-E_F$ & split 4 & $1.35$ & $10.90-E_F$ \\
\end{tabular}
\end{ruledtabular}
\footnotetext[1]{Ref.~\onlinecite{wright02}.}
\end{table*}
%

\subsubsection{$\mathrm{C_N-V_{Ga}}$}

Both gallium and nitrogen vacancies were historically well investigated
as parts of native defects in GaN. Earlier theoretical studies are based on the standard
DFT (LDA and GGA)~\cite{neugebauer94,mattila97,ganchenkova06}. Recently hybrid functionals
are used increasingly~\cite{yan12,gillen13,lyons15}. Here we studied the complexes made
of substitutional carbon and vacancy. In this subsection the results for $\mathrm{C_N-V_{Ga}}$
are given. Then the results of $\mathrm{C_{Ga}-V_N}$ will be presented in the following
subsection.

\begin{figure}
\includegraphics[width=4cm]{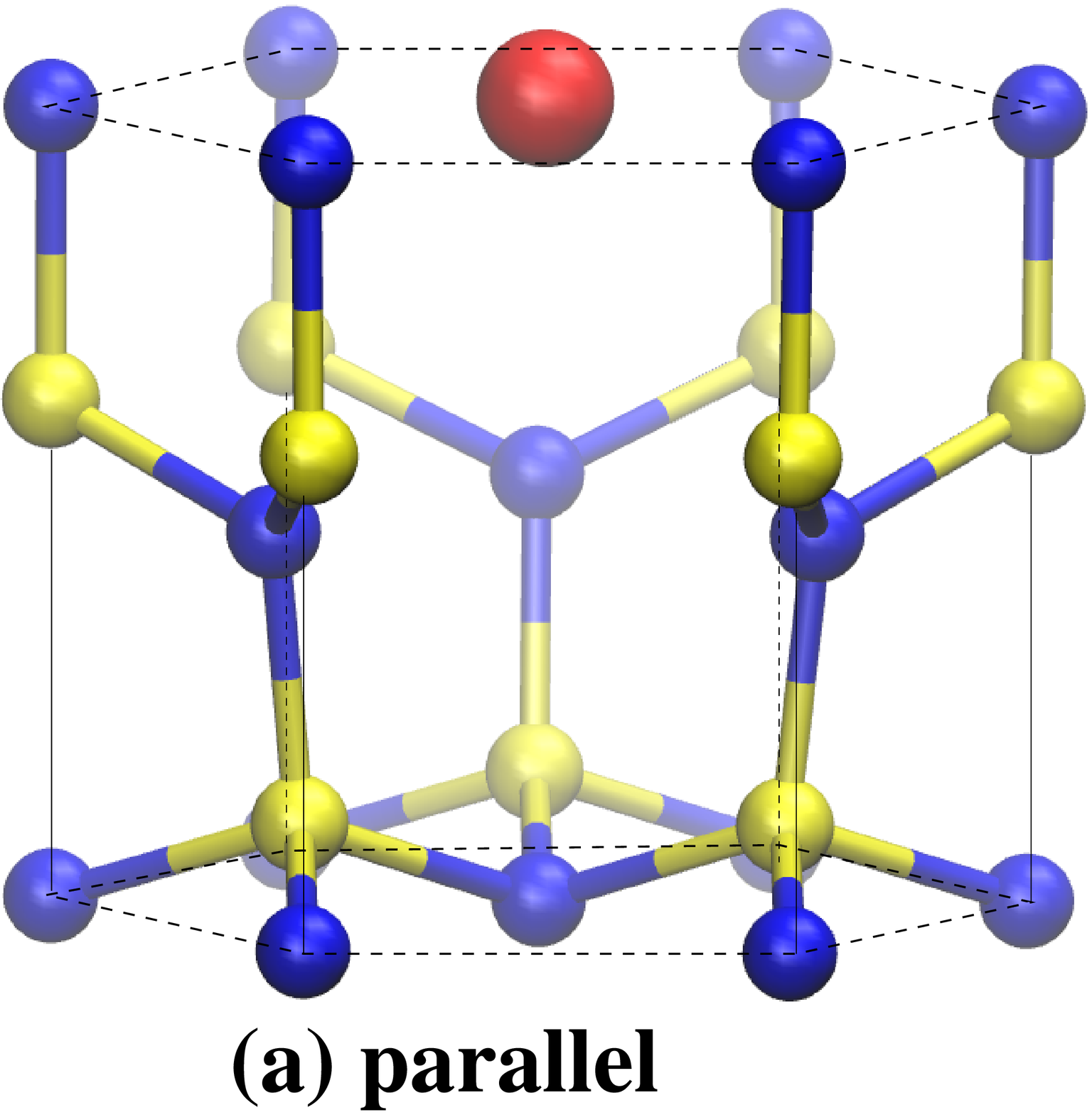}%
\includegraphics[width=4cm]{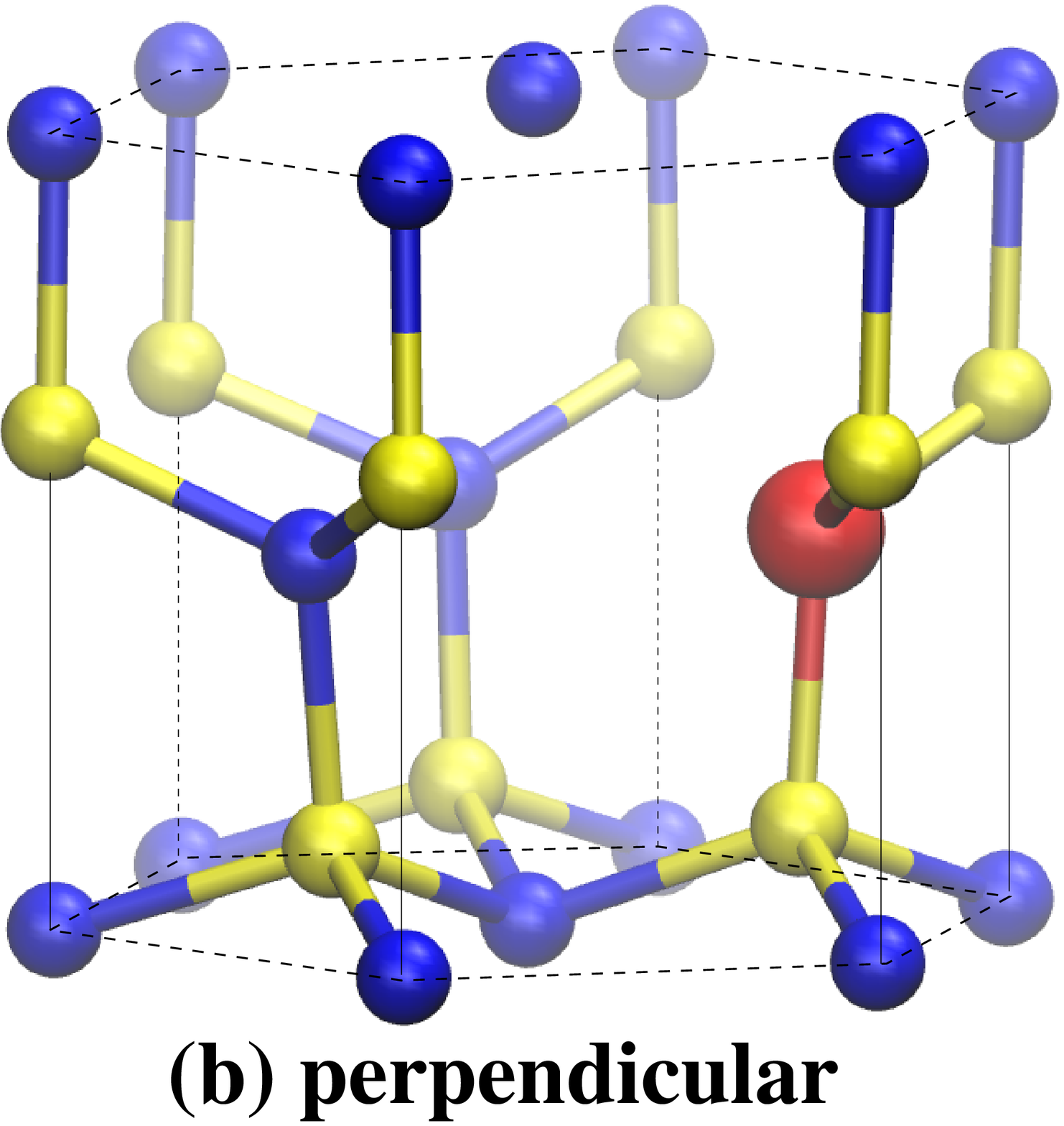}%
\caption{\label{f:geoCNVGa}Ball and stick representations of $\mathrm{C_N-V_{Ga}}$ complex.
$\mathrm{V_{Ga}}$ is located at the center of hexagonal prism.
(a) $\mathrm{C_N}$ and $\mathrm{V_{Ga}}$ are located parallel to the $c$-axis.
(b) $\mathrm{C_N}$ and $\mathrm{V_{Ga}}$ are located perpendicular to the $c$-axis.
}
\end{figure}

Two different configurations are found to be favorable for this
complex. In one configuration, $\mathrm{C_N}$ and $\mathrm{V_{Ga}}$
are located parallel to the $c$-axis, as shown in
Fig.~\ref{f:geoCNVGa}(a). In the other configuration, $\mathrm{C_N}$
and $\mathrm{V_{Ga}}$ are located perpendicular to the $c$-axis, as
depicted in Fig.~\ref{f:geoCNVGa}(b). Both configurations have very
similar formation energies, with less than 0.05\,eV difference, but
the parallel configuration possesses slightly lower formation
energies.

Formation energy values obtained for the $\mathrm{C_N-V_{Ga}}$
complex are plotted in Fig.~\ref{f:EfCc} with brown solid line. The
1+, 0 (neutral) and $1-$, $2-$ and $3-$ charge states are present in
the band gap. As opposed to previous LDA
calculations~\cite{wright02} in which this complex was found to
behaves only as a deep acceptor, the present result indicates that
$\mathrm{C_N-V_{Ga}}$ shows amphoteric behavior. The (+/0) donor
level appears at 0.61\,eV above the valence band edge, while the
(0/$-$), ($-$/$2-$) and ($2-$/$3-$) acceptor levels are at 1.70,
1.97 and 2.29\,eV, respectively. Examining the binding energy of
this complex, from Fig.~\ref{f:EB} it can be seen that at 2.14\,eV
above the valence band edge the binding energy becomes negative and
the complex can no longer be stable.
In addition, in $n$-type GaN (upper half of the band gap), both $\mathrm{C_N}$
and $\mathrm{V_{Ga}}$ are negatively charged and are expected to repel each other.
This may impede the formation of this complex in $n$-type GaN.
The C--C bond lengths and formation energies in each charge state are summarized
in Table~\ref{t:CNVGa} together with previous LDA results~\cite{wright02} for
comparison.

%

%
\begin{table*}
\caption{Comparison of structural and electronic properties of $\mathrm{C_N-V_{Ga}}$ in the different charge states $q$ between the previous result in LDA~\cite{wright02} and this work in HSE. The preferred forms with lowest energies in each charge state, the C--$\mathrm{V_{Ga}}$ distances ($d_{\mathrm{C-V}}$ in \AA), where the $\mathrm{V_{Ga}}$ position is taken to be the center of mass for the six surrounding Ga atoms, and the formation energies ($E_f$ in eV) are presented.\label{t:CNVGa}}
\begin{ruledtabular}
\begin{tabular}{ccccccc}
 & \multicolumn{3}{c}{LDA~\footnotemark[1]} & \multicolumn{3}{c}{HSE (this work)} \\ \hline
 $q$  & form & $d_{\mathrm{C-V}}$ & $E_f$ & form & $d_{\mathrm{C-V}}$ & $E_f$ \\
 $1+$ & --                  & -- & --           & $\parallel$ & $2.21$ & $8.99+E_F$ \\
 $0$  & $\parallel$/$\perp$ & -- & $9.85$       & $\parallel$ & $2.00$ & $9.60$ \\
 $1-$ & $\perp$             & -- & $10.11-E_F$  & $\parallel$ & $2.05$ & $11.30-E_F$ \\
 $2-$ & $\perp$             & -- & $10.73-2E_F$ & $\parallel$ & $2.08$ & $13.27-2E_F$ \\
 $3-$ & $\perp$             & -- & $12.09-3E_F$ & $\parallel$ & $2.12$ & $15.57-3E_F$ \\
 $4-$ & $\parallel$         & -- & $14.17-4E_F$ & $\parallel$ & $2.01$ & $20.06-4E_F$ \\
\end{tabular}
\end{ruledtabular}
\footnotetext[1]{Ref.~\onlinecite{wright02}.}
\end{table*}
%

\subsubsection{$\mathrm{C_{Ga}-V_N}$}

\begin{figure}
\includegraphics[width=4cm]{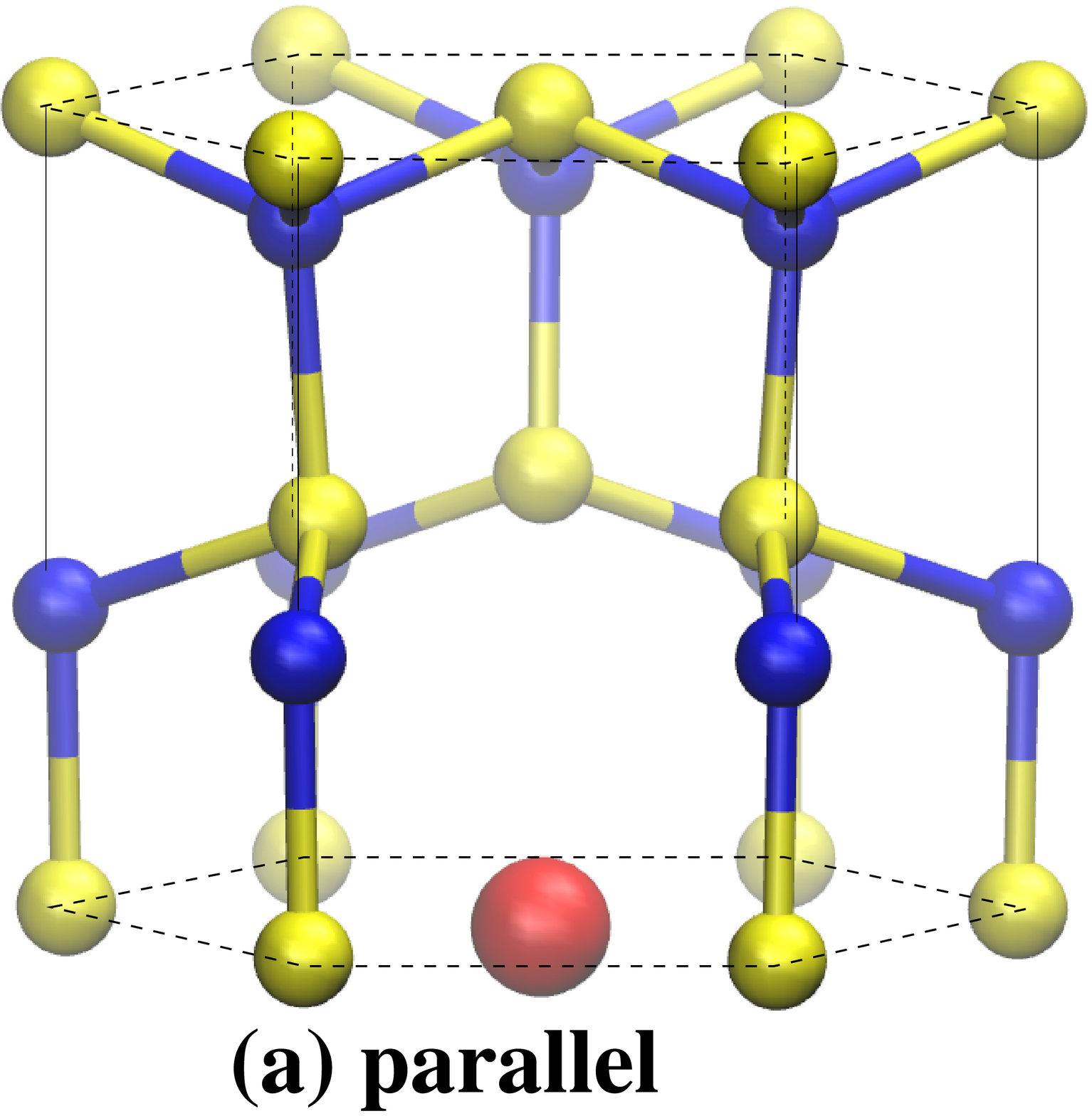}%
\includegraphics[width=4cm]{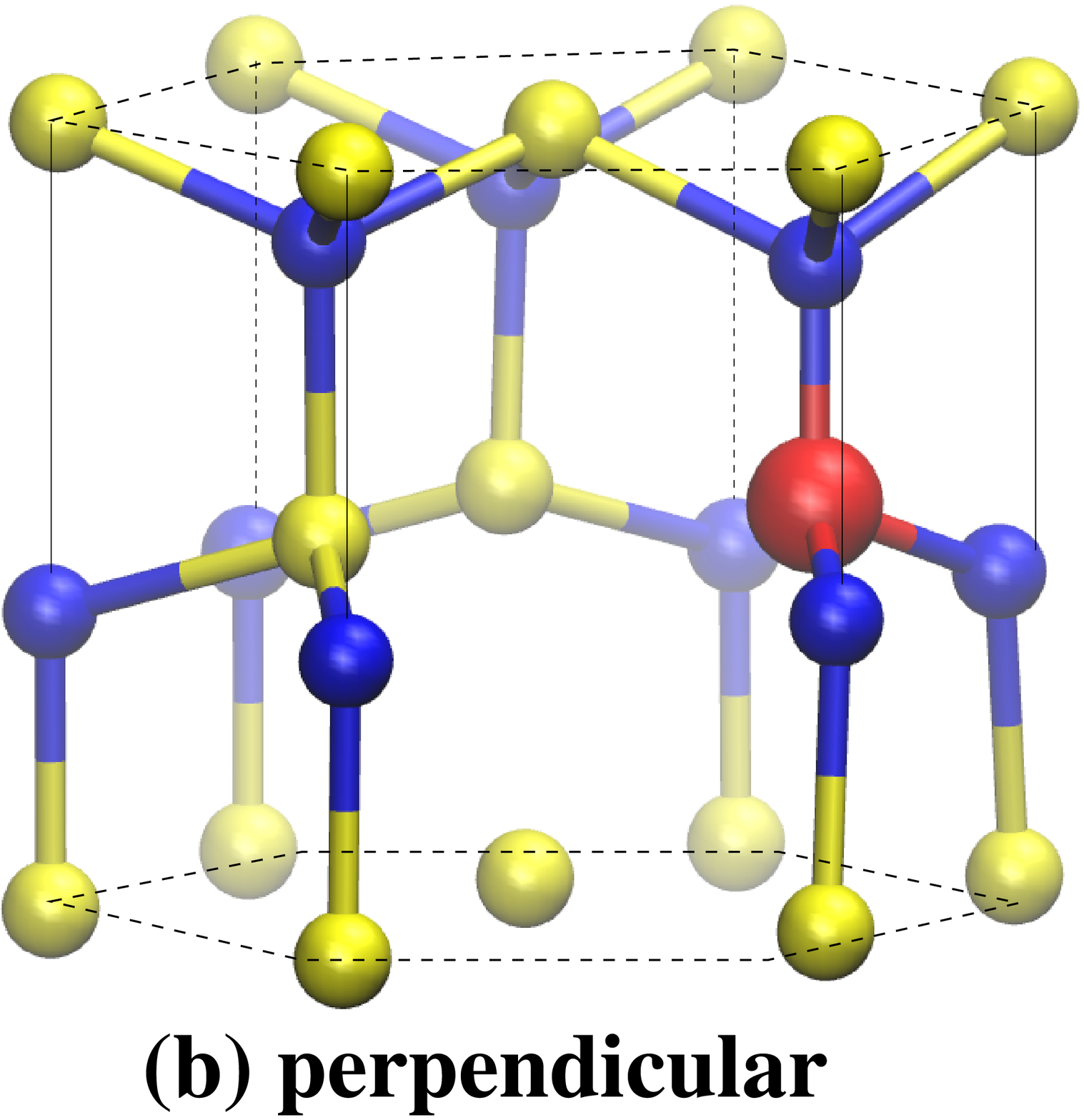}%
\caption{\label{f:geoCGaVN}Ball and stick representations of $\mathrm{C_{Ga}-V_N}$ complex.
$\mathrm{V_N}$ is located at the center of hexagonal prism.
(a) $\mathrm{C_{Ga}}$ and $\mathrm{V_N}$ are located parallel to the $c$-axis.
(b) $\mathrm{C_{Ga}}$ and $\mathrm{V_N}$ are located perpendicular to the $c$-axis.
}
\end{figure}

Similarly to the previous case, the $\mathrm{C_{Ga}-V_N}$ complex is
also found to assume parallel and perpendicular configurations. They
are presented in Figs.~\ref{f:geoCGaVN}(a) and (b), respectively. In
the parallel configuration, $\mathrm{C_{Ga}}$ and $\mathrm{V_N}$ are
located parallel to the $c$-axis and in the perpendicular
configuration, $\mathrm{C_{Ga}}$ and $\mathrm{V_N}$ are located
perpendicular to the $c$-axis. Once again, both configurations have
very similar formation energies. In the 4+ and 2+ charge states, the
perpendicular configuration is slightly more stable than the
parallel configuration, whereas in the 1+ and the 0 (neutral) charge states,
the parallel configuration is slightly more stable than the
perpendicular configuration.

Formation energy values obtained for $\mathrm{C_{Ga}-V_N}$ complex
are plotted in Fig.~\ref{f:EfCc} with orange dashed line. The
4+, 2+, 1+ and 0 (neutral) charge states area present in the band
gap. Therefore, $\mathrm{C_{Ga}-V_N}$ complex behaves as a potential
donor. The 4+ charge state is favorable up to 0.47\,eV above the
valence band edge.
Subsequently the 2+ charge state up to 2.40\,eV
and the the 1+ charge state up to 3.23\,eV become favorable. Finally the 0 (neutral)
charge state becomes the most favorable above 3.23\,eV. The
binding energy of this complex, shown in Fig.~\ref{f:EB}, is at
least 0.92\,eV (close to the VBM) and eventually increases to
2\,eV (close to CBM). Consequently, the
$\mathrm{C_{Ga}-V_N}$ complex is stable for all values of Fermi
energy within the band gap.
However, in $p$-type GaN, both $\mathrm{C_{Ga}}$ and $\mathrm{V_N}$ are
positively charged and are expected to repel each other. Thus, in $p$-type GaN
the formation of this complex may be impeded.
The C--V distances and formation energies in each charge state are summarized
in Table~\ref{t:CGaVN}.


%
%

%
\begin{table*}
\caption{The preferred forms with lowest energies in each charge state, the C--$\mathrm{V_{N}}$ distances ($d_{\mathrm{C-V}}$ in \AA), where the $\mathrm{V_{N}}$ position is taken to be the center of mass for the six surrounding N atoms, and the formation energies ($E_f$ in eV) are presented.\label{t:CGaVN}}
\begin{ruledtabular}
\begin{tabular}{cccc}
 & \multicolumn{3}{c}{HSE (this work)} \\ \hline
 $q$  & form & $d_{\mathrm{C-V}}$ & $E_f$ \\
 $4+$ & $\perp$     & $2.52$ & $0.90+4E_F$ \\
 $2+$ & $\perp$     & $2.54$ & $1.84+2E_F$ \\
 $1+$ & $\parallel$ & $2.30$ & $4.24+E_F$ \\
 $0$  & $\parallel$ & $2.07$ & $7.47$ \\
\end{tabular}
\end{ruledtabular}
\end{table*}
%

\section{Discussion\label{s:discussion}}

This section presents a comparison between the experimental data
available in the literature and the calculated results that have
been outlined in the previous section. In general we expect that,
based on Eq.~(\ref{eq:conc}), defects and complexes with lower
formation energies may be present in higher concentrations and be
the dominant carbon forms. However, this does not exclude the
existence of other forms of the impurities. For example, in the case
of $n$-type material, in which the Fermi energy is located in the
upper half of the band gap, $\mathrm{C_N}$ with $-1$ charge state is
expected to have the lowest formation energy among all kinds of
C-related defects considered here. Therefore, $\mathrm{C_N}$ is
expected to be the dominant form of carbon inclusion. However,
recent experimental results suggest that, along with $\mathrm{C_N}$,
other form of carbon are present in bulk GaN~\cite{tompkins11}.
Among the carbon--carbon and carbon--vacancy complexes studied
in this manuscript, $\mathrm{C_N-C_{Ga}}$ and $\mathrm{C_N-V_{Ga}}$
have the lowest formation energies in $n$-type GaN. Therefore they are
also possible candidates for the carbon related defects commonly observed in GaN,
although their formation energies are still much higher than that of $\mathrm{C_N}$.
However, we avoid for the moment making any statement about
the possible dominant type of carbon inclusions and simply compare
the calculated trap level with the measured one and try to identify
which of the carbon-related defects or complexes may be responsible
for it.
%
Detailed analysis considering the impurity concentration
will be done in the subsequent paper~\cite{Matsubara_C2}, after examining all types of carbon
related complexes.

As already pointed out in Section~\ref{s:intro}, a number of
experiments have been performed to try to identify carbon-related
traps in GaN. Table~\ref{t:ExpLv} summarizes the experimental data
available in the literature.
\begin{table*}
\caption{Experimentally observed C-related trap levels. Activation
energies obtained by thermal methods such as DLTS and by optical
methods such DLOS are denoted by $E_{\mathrm{TH}}$ and
$E_{\mathrm{OPT}}$, respectively (in eV).\label{t:ExpLv}}
\begin{ruledtabular}
\begin{tabular}{lllll}
\multicolumn{2}{c}{Armstrong \emph{et al.}\footnotemark[1]} & Shah \emph{et al.}\footnotemark[2] & Polyakov \emph{et al.}\footnotemark[3] & Honda \emph{et al.}\footnotemark[4] \\ \hline
$E_c - E_{\mathrm{OPT}}$ & $E_c - E_{\mathrm{TH}}$/$E_v + E_{\mathrm{TH}}$ & $E_c - E_{\mathrm{TH}}$ & $E_c - E_{\mathrm{OPT}}$ & $E_c - E_{\mathrm{TH}}$/$E_v + E_{\mathrm{TH}}$ \\ \hline
$E_c - 1.35$ & $E_c - 0.11$ & $E_c - 0.13$ & $E_c - 1.3/1.4$ & $E_c - 0.40$ \\
$E_c - 3.0$ & $E_v + 0.9$ & $E_c - 2.69$ & $E_c - 2.7/2.8$ & $E_v + 0.86$ \\
$E_c - 3.28$ &              & $E_c - 3.20$ & $E_c - 3$ & \\
$E_c -1.94/2.05$ & & & & \\
\end{tabular}
\end{ruledtabular}
\footnotemark[1]{Ref.~\onlinecite{armstrong05}.}
\footnotemark[2]{Ref.~\onlinecite{shah12}.}
\footnotemark[3]{Ref.~\onlinecite{polyakov13}.}
\footnotemark[4]{Ref.~\onlinecite{honda12}.}
\end{table*}

\begin{figure}
\includegraphics[width=\columnwidth]{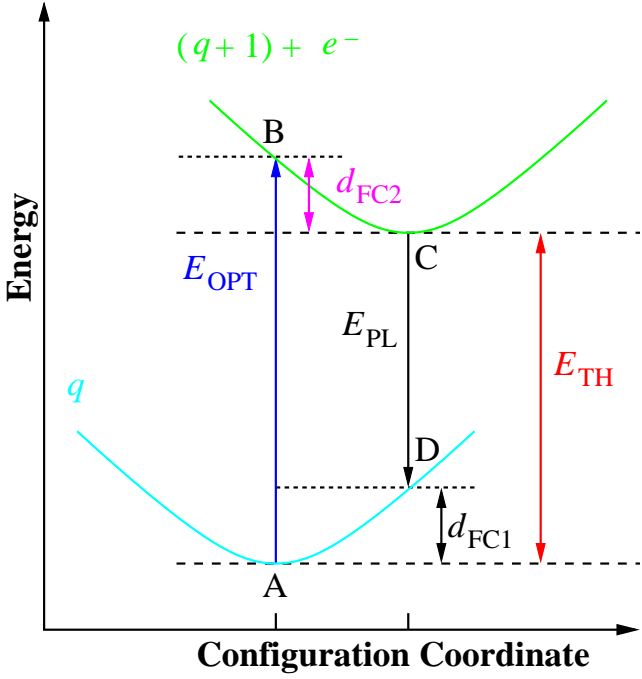}%
\caption{\label{f:FC}Schematic configuration coordinate diagram of the
electron capture process to show
the relationship between thermal activation energy ($E_{\mathrm{TH}}$)
and optical activation energy ($E_{\mathrm{OPT}}$).}
\end{figure}

Activation energies are experimentally determined using two main
techniques: DLTS and DLOS that use thermal ionization and optical
ionization of traps respectively. DLTS provides information on the
thermal activation energy ($E_{\mathrm{TH}}$), whereas DLOS on the
optical activation energy ($E_{\mathrm{OPT}}$). The difference
between $E_{\mathrm{TH}}$ and $E_{\mathrm{OPT}}$ is schematically
depicted in Fig.~\ref{f:FC}, where only the electron capture process
is described~\cite{holecapture}.
Specifically, $d_{\mathrm{FC1}}$ and $d_{\mathrm{FC2}}$ are the
Franck-Condon shifts, which are the energies transferred to the
lattice due to the relaxation process between the two equilibrium
configurations in the respective charge states. Furthermore, from
Fig.~\ref{f:FC} we find that $E_{\mathrm{OPT}} = E_{\mathrm{TH}} +
d_{\mathrm{FC2}}$.

%


Armstrong~\cite{armstrong05} and coworkers investigated five trap
levels that were determined to be C-related. Among these, a trap
observed at $E_c - 3.0$\,eV by DLOS and another trap at $E_v + 0.9$
eV by DLTS were considered to have the same origin. Additionally,
trap levels at $E_c -1.35$\,eV, $E_c - 3.28$\,eV and $E_c -
0.11$\,eV were assigned to $\mathrm{C_I}$ (0/2$-$), $\mathrm{C_N}$
(0/$-$) and $\mathrm{C_{Ga}}$ (+/0), respectively. The origins of
the other two levels, $E_c - 3.0/E_v +0.9$\,eV and $E_c -
1.94/2.05$\,eV, were not specified. The results obtained by Shah
\emph{et al.}~\cite{shah12} point to three trap levels. One of them
located at $E_c - 0.13$\,eV was assigned to $\mathrm{C_{Ga}}$ (+/0).
The trap levels at $E_c - 2.69$\,eV and $E_c -3.20$\,eV were assumed
to be related to $\mathrm{C_N}$. The latter was assigned to
$\mathrm{C_N}$ (0/$-$), whereas the former to a $\mathrm{C_N}$
related complex or gallium vacancies. Using room temperature
photocurrent measurement, Polyakov \emph{et al.}\ found three optical
thresholds corresponding to $E_c - 1.3/1.4$\,eV, $E_c - 2.7/2.8$\,eV
and $E_c -3$\,eV. The first one was attributed to $\mathrm{C_I}$
(0/2$-$) level, while the other two levels were not assigned. The
assignments of the trap levels to specific carbon related defects
performed by three experimental groups considered above were carried
out on the basis of existing LDA results. Finally, Honda and
coworkers~\cite{honda12} using DLTS and MCTS observed three
C-related trap levels. Unfortunately, energy levels for only two of
them were reported. One at $E_c - 0.40$\,eV and the other at $E_v +
0.86$\,eV. Using HSE calculated energy~\cite{lyons10}, $E_v + 0.86$\,eV was
assigned to $\mathrm{C_N}$ (0/$-$), while the origin of the trap at
$E_c - 0.40$\,eV was not specified.

The formation energy values presented in Sec.\,\ref{ss:singleC}
and Sec.\,\ref{ss:complex} were computed under the assumption of
thermodynamic equilibrium. As a result, the transition between the
levels of two charge states ($q$/$q^{\prime}$) is characterized
using a quantity called thermal ionization energies. With reference
to Fig.~\ref{f:FC}, we define $E_{\mathrm{TH}}$ as the energy
difference between configurations C and A, $E_{\mathrm{C}} -
E_{\mathrm{A}}$ that is defined as $E_f^{q+1} + E_g - E_f^q$, where
$E_g$ is the energy, corresponding to the band gap, necessary to
add an electron to the conduction band. Furthermore,
$E_{\mathrm{TH}}$ can also be expressed as $E_{\mathrm{TH}} = E_g -
\varepsilon(q/q+1)$, using Eq.~(\ref{eq:eqq}). Similarly, from
Fig.~\ref{f:FC} $E_{\mathrm{OPT}}$ is defined as the energy
difference between the configurations B and A or $E_{\mathrm{B}} -
E_{\mathrm{A}}$. Specifically, $E_{\mathrm{B}}$ is calculated
starting from the formation energy of the state having the geometry
of A with a charge state $q+1$ and adding $E_g$ to account for the
energy necessary to promote an electron to the conduction band.

Using the previously derived relationships, both thermal and optical
activation energies obtained by our calculations are summarized in
Tables~\ref{t:CalcTh1} for single impurity carbon,
and~\ref{t:CalcTh2} for carbon complexes. We point out that in these
tables we also report transition levels involving thermodynamically
unstable charge states: (4+/3+) and (3+/2+) of $\mathrm{C_I}$,
(3+/2+) and (2+/+) of $\mathrm{C_I-C_N}$, (4+/3+) and (3+/2+) of
$\mathrm{C_{Ga}-V_N}$, which are experimentally accessible using
dynamical techniques such as DLTS and DLOS.
Furthermore, values that are reported in bold
and enclosed in a box are the proposed assignment emerging from our
numerical analysis.


%
\begin{table*}
\caption{Thermodynamic transition levels [$\varepsilon(q/q^{\prime})$], thermal activation energies ($E_{\mathrm{TH}}$)
and optical activation energies ($E_{\mathrm{OPT}}$) obtained from our calculated
results for single impurity of C. The levels which do not appear within the band gap are denoted as horizontal bar. The energy levels close to the experimental ones are denoted in bold.\label{t:CalcTh1}}
\begin{ruledtabular}
\begin{tabular}{lllll}
Form & ($q/q^{\prime}$) & $\varepsilon(q/q^{\prime})$ & $E_{\mathrm{TH}}$ & $E_{\mathrm{OP
T}}$ \\ \hline
$\mathrm{C_{Ga}}$ & (+/0) & -- & -- & -- \\
                 & (0/$-$) & -- & -- & -- \\
$\mathrm{C_N}$ & (+/0) & 0.25 & \fbox{\textbf{3.20}} & -- \\
               & (0/$-$) & \fbox{\textbf{0.89}} & \fbox{\textbf{2.56}} & \fbox{\textbf{2.91}} \\
$\mathrm{C_I}$ & (4+/2+) & \fbox{\textbf{0.81}} & \fbox{\textbf{2.64}} & -- \\
               & (4+/3+) & 2.63 & 0.82 & \fbox{\textbf{3.11}} \\
               & (3+/2+) & -- & -- & -- \\
               & (2+/+) & 1.62 & 1.83 & \fbox{\textbf{2.87}} \\
               & (+/0) & 2.58 & 0.87 & 1.58 \\
               & (0/$-$) & 3.20 & \fbox{\textbf{0.25}} & 0.82 \\
\end{tabular}
\end{ruledtabular}
\end{table*}
\begin{table*}
\caption{Thermodynamic transition levels [$\varepsilon(q/q^{\prime})$], thermal activation energies ($E_{\mathrm{TH}}$)
and optical activation energies ($E_{\mathrm{OPT}}$) obtained from our calculated
results for C complexes. The levels which do not appear within the band gap are denoted as horizontal bar. The energy levels close to the experimental ones are denoted in bold.\label{t:CalcTh2}}
\begin{ruledtabular}
\begin{tabular}{lllll}
Form & ($q/q^{\prime}$) & $\varepsilon(q/q^{\prime})$ & $E_{\mathrm{TH}}$ & $E_{\mathrm{OP
T}}$ \\ \hline
$\mathrm{C_N-C_{Ga}}$ & (2+/+) & 0.05 & 3.40 & -- \\
                      & (+/0) & 0.52 & 2.93 & -- \\
$\mathrm{C_I-C_N}$ & (3+/+) & \fbox{\textbf{0.88}} & \fbox{\textbf{2.57}} & -- \\
                   & (3+/2+) & 2.28 & 1.17 & -- \\
                   & (2+/+) & -- & -- & -- \\
                   & (+/0) & 3.21 & 0.24 & 1.03 \\
$\mathrm{C_I-C_{Ga}}$ & (3+/2+) & 1.79 & 1.66 & \fbox{\textbf{2.79}} \\
                      & (2+/+) & 2.27 & 1.18 & \fbox{\textbf{1.96}} \\
                      & (+/0) & 2.60 & 0.85 & \fbox{\textbf{1.47}} \\
                      & (0/$-$) & 2.84 & 0.61 & \fbox{\textbf{1.31}} \\
$\mathrm{C_N-V_{Ga}}$ & (+/0) & 0.61 & 2.84 & \fbox{\textbf{3.21}} \\
                      & (0/$-$) & 1.70 & 1.75 & 2.08 \\
                      & ($-$/2$-$) & 1.97 & 1.48 & 1.49 \\
                      & (2$-$/3$-$) & 2.29 & 1.16 & 1.27 \\
$\mathrm{C_{Ga}-V_N}$ & (4+/2+) & 0.47 & 2.98 & -- \\
                      & (4+/3+) & 0.55 & 2.90 & -- \\
                      & (3+/2+) & 0.40 & 3.05 & -- \\
                      & (2+/+) & 2.40 & 1.05 & \fbox{\textbf{2.13}} \\
                      & (+/0) & 3.23 & \fbox{\textbf{0.22}} & 1.06 \\
\end{tabular}
\end{ruledtabular}
\end{table*}

We consider first the trap level located at $E_c - 3.0/E_v +
0.9$\,eV obtained by Armstrong \emph{et al.}\ and that was also
observed by Polyakov\,\emph{et al.}\ as well as by Honda\,\emph{et
al.}. While Armstrong\,\emph{et al.}\ and Polyakov\,\emph{et al.}\ did
not give a clear assignment to this trap level, Honda\,\emph{et
al.}~\cite{honda12} assigned it to $\mathrm{C_N}$ (0/$-$) by
comparing the experimental value to the calculated result obtained
using HSE by Lyons\,\emph{et al.}~\cite{lyons10}. Indeed our
calculated results support this assignment. In our case, the
$\mathrm{C_N}$ (0/$-$) level is located at 0.89\,eV above valence
band maximum with 2.91\,eV optical activation energy
($E_{\mathrm{OPT}}$) as shown in Table~\ref{t:CalcTh1}. Furthermore,
the thermal activation energy of this trap is calculated to be
$E_{\mathrm{TH}}$\,=\,2.56\,eV. This value is in good agreement with
the value of 2.60\,eV reported in Ref.~\onlinecite{lyons14}, where its assignment
was done for the onset near 2.5\,eV by the photoluminescence excitation data~\cite{ogino80}.
In addition to the $\mathrm{C_N}$ (0/$-$) level, $\mathrm{C_I}$ (4+/2+)
level is located at 0.81\,eV above the VBM, which corresponds to the 2.64\,eV
thermal activation energy (Table~\ref{t:CalcTh1}).
Related optical activation energy, $\mathrm{C_I}$ (4+/3+)
is calculated as 3.11\,eV, which is close to the experimental value $E_c - 3.0$ eV.
Similarly, the (3+/+) level of $\mathrm{C_I-C_N}$, is positioned at
0.88\,eV above the VBM with $E_{\mathrm{TH}}=2.57$\,eV (Table~\ref{t:CalcTh2}).
Consequently we argue that the trap
level at $E_c - 2.69$\,eV, originally measured by Shah\,\emph{et al.}~\cite{shah12}
and that was considered to be related to
$\mathrm{C_N}$ complex or gallium vacancy, is in reality due to
$\mathrm{C_N}$ (0/$-$)
with possible contributions from $\mathrm{C_I}$ and $\mathrm{C_I-C_N}$.


We turn our attention to the trap levels originally assigned to
$\mathrm{C_N}$ (0/$-$), specifically the one located at $E_c -
3.28$\,eV measured by Armstrong\,\emph{et al.}\ and $E_c - 3.20$\,eV
level by Shah\,\emph{et al.}. Since Shah\,\emph{et al.}\ measured this
energy with MCTS we assume that this is a thermal activation energy.
From Tables~\ref{t:CalcTh1} and~\ref{t:CalcTh2} we can observe that
the energies corresponding to (+/0) of $\mathrm{C_N}$ gives the value of
3.20\,eV.
On the other hand, we notice that the trap level observed by
Armstrong\,\emph{et al.}\ with DLOS should be treated as an optical
activation energy. In this case we observe that
the (+/0) of $\mathrm{C_N-V_{Ga}}$
(3.21\,eV) is calculated as an optical trap level, which has the energy
very close to experimental one (3.28\,eV).

The next one is the trap level at $E_c - 1.35$\,eV observed by
Armstrong\,\emph{et al.}\ and Polyakov\,\emph{et al.}\ that was
assigned to $\mathrm{C_I}$ (0/$2-$) based on a 1.13\,eV thermal
activation energy computed by LDA~\cite{wright02}. First we
point out that, based on our HSE calculation, the (0/$2-$) state
is energetically unfavorable while the (0/$-$), that was found to be
unfavorable with LDA, is now possible. For this (0/$-$) state we
compute an optical activation energy of 0.82\,eV that corresponds to
a thermal activation energy of 0.25\,eV. Consequently,
$\mathrm{C_I}$ (0/$-$) cannot be the origin of the level $E_c -
1.35$\,eV observed by Armstrong\,\emph{et al.} and Polyakov\,\emph{et al}.
Additionally, we
notice that the calculated (+/0) $\mathrm{C_I}$ level from HSE
results in a optical activation energy of 1.58\,eV, corresponding to
a thermal activation energy of 0.87\,eV. Therefore based on our
results, the origin of $E_c - 1.35$\,eV level is not likely to be
related to any of the $\mathrm{C_I}$ states. Finally, based on
HSE the (0/$-$) and (+/0)
states of the $\mathrm{C_I - C_{Ga}}$ complex have
a 1.31\,eV and 1.47\,eV optical activation energies, respectively.
As a result, $\mathrm{C_I - C_{Ga}}$ is likely to
be the origin of the experimentally observed $E_c - 1.35$\,eV level.
Note that the ($-/2-$) and ($2-$/$3-$) states of the $\mathrm{C_N - V_{Ga}}$
complex have $E_{\mathrm{OPT}}=1.27$\,eV and 1.49\,eV, respectively,
but this complex is unstable
in this energy region with negative or barely positive binding energies (Fig.~\ref{f:EB}).


Honda\,\emph{et al.}\ observed, using DLTS, a $E_c - 0.40$\,eV level
and concluded that it was C-related, but did not mention its origin.
Based on our calculations the (0/$-$) level of $\mathrm{C_I}$ has a
0.25\,eV thermal activation energy. Furthermore, the theoretical
values of the activation energy for the (+/0) level of
$\mathrm{C_{Ga}-V_N}$ is {0.22}\,eV. Therefore, it is likely that the
origin of the $E_c - 0.40$\,eV level is the (0/$-$) state of
$\mathrm{C_I}$ and/or the (+/0) state of $\mathrm{C_{Ga}-V_N}$.
The (+/0) level of $\mathrm{C_I-C_N}$ has $E_{\mathrm{TH}}=0.24$\,eV.
However, at this energy range this complex is unlikely to form due to the
barely positive binding energy.

The $E_c -1.94/2.05$\,eV level observed with DLOS by Armstrong
\emph{et al.}\ was assumed to be C-related but its physical form
remained unknown. From our calculated results two levels have
similar optical activation energies. One is (2+/+) level of
$\mathrm{C_I - C_{Ga}}$ with 1.96\,eV and the other is (2+/+) level
of $\mathrm{C_{Ga}-V_N}$ with 2.13\,eV.
These two states are the likely candidates to explain the physical origin of the $E_c -
1.94/2.05$\,eV trap level.
The ($0/-$) level of weekly bounded $\mathrm{C_N-V_{Ga}}$ with $E_{\mathrm{OPT}}=2.08$\,eV
is unlikely to contribute.


Polyakov \emph{et al.}\ observed a optical threshold at 2.7--2.8\,eV
by photocurrent spectra measurement. The origin of this trap level
is likely to be the (2+/+) level of $\mathrm{C_I}$
and/or the (3+/2+) level of $\mathrm{C_I-C_{Ga}}$,
whose optical activation energies are computed to be 2.87\,eV
and 2.79\,eV, respectively.

The last two energy levels we consider are the one located at $E_c -
0.11$\,eV measured by Armstrong\,\emph{et al.}\ and at $E_c - 0.13$\,eV
measured by Shah\,\emph{et al.}. These two levels  were assigned to
$\mathrm{C_{Ga}}$ based on LDA results~\cite{wright02}, for which
the (+/0) transition level was computed to be about 0.2\,eV below
CBM. However, our HSE calculation shows that such (+/0) level of
$\mathrm{C_{Ga}}$ does not appear within the band gap, but it is in
the conduction band approximately 0.5\,eV above the CBM. Recent HSE
calculation by Lyons\,\emph{et al.}~\cite{lyons14} does not show
such level within the band gap either. From our results we evince
that the (+/0) level of $\mathrm{C_I - C_N}$ complex has a 0.24\,eV
thermal activation energy. Unfortunately this complex is unlikely to
form when the Fermi energy is around this value because its biding
energy is barely positive as we can see from Fig.~\ref{f:EB}.
Based on our calculations, the most likely candidate is the (+/0) state of
$\mathrm{C_{Ga}-V_N}$ located at 0.22\,eV below the CBM and/or the (0/$-$) state
of $\mathrm{C_I}$ at 0.25\,eV below the CBM. Therefore, it
is probable that these levels have the same origin of the $E_c
-0.40$\,eV level observed by Honda \emph{et al.}.

Finally, we also calculated vibrational frequencies of relevant charge states
in the cases of $\mathrm{C_I}$, $\mathrm{C_I-C_{Ga}}$ and $\mathrm{C_I-C_{N}}$ complexes.
Our calculations are based on the finite difference method also implemented in the
VASP code, where small (both positive and negative by 0.015\,\AA) displacements are introduced.
These local vibrational
modes may provide alternative information for the experimental detection of these
carbon related impurities.
The frequencies for the breathing modes in the $4+$ and $3+$ charge states of $\mathrm{C_I}$
and for the stretching modes in the other charge states of $\mathrm{C_I}$ and all the charge
states of $\mathrm{C_I-C_N}$ and $\mathrm{C_I-C_{Ga}}$ are summarized in Table~\ref{t:Vib}.

\begin{table}
\caption{Vibrational frequencies for $\mathrm{C_I}$, $\mathrm{C_I-C_{Ga}}$ and
$\mathrm{C_I-C_N}$ (in cm$^{-1}$).
The breathing modes are given in the $4+$ and $3+$ charge states of $\mathrm{C_I}$ and the stretching
modes are given in the other charge states of $\mathrm{C_I}$ and all the charge states of
$\mathrm{C_I-C_{Ga}}$ and $\mathrm{C_I-C_N}$.\label{t:Vib}}
\begin{ruledtabular}
\begin{tabular}{lll}
Form & $q$ & vibrational frequency \\ \hline
$\mathrm{C_I}$ & $4+$ & 1004 \\
               & $2+$ & 2213 \\
               & $1+$ & 1839 \\
               & $0$  & 1546 \\
               & $1-$ & 1279 \\
$\mathrm{C_I-C_{Ga}}$ & $3+$ & 1455 \\
                      & $2+$ & 1462 \\
                      & $1+$ & 1459 \\
                      & $0$ & 1090 \\
                      & $1-$ & 1574 \\
$\mathrm{C_I-C_N}$ & $3+$ & 1246 \\
                   & $1+$ & 2049 \\
\end{tabular}
\end{ruledtabular}
\end{table}

\section{Conclusion\label{s:conclusion}}

We have performed first-principles calculations using HSE hybrid
density functional in the framework of DFT to investigate the
characteristics of various forms of carbon inclusions in GaN. We
have considered single impurities, $\mathrm{C_N}$, $\mathrm{C_{Ga}}$
and $\mathrm{C_I}$, as well as their complexes
$\mathrm{C_N-C_{Ga}}$, $\mathrm{C_I-C_N}$, $\mathrm{C_I-C_{Ga}}$,
$\mathrm{C_N-V_{Ga}}$ and $\mathrm{C_{Ga}-V_N}$. For all these
configurations, different charge states have been considered and
their geometries are fully optimized. Formation and binding energies
of complexes have been computed and thermodynamic transition levels are
obtained.

Among single impurities, $\mathrm{C_N}$ behaves mainly as a deep
acceptor, $\mathrm{C_{Ga}}$ acts as a donor without inducing states
in the band gap and $\mathrm{C_I}$ shows amphoteric behavior. Both
in N-rich and Ga-rich conditions the 4+ charge state of
$\mathrm{C_I}$, which assumes an octahedral interstitial position,
is favorable close to the VBM ($p$-type region). The $1-$ charge
state of $\mathrm{C_N}$ is favorable close to the CBM ($n$-type
region), and $\mathrm{C_{Ga}}$ is also favorable in $p$-type region
but only in N-rich conditions.

Complexes made of combinations of single impurities, specifically
$\mathrm{C_N-C_{Ga}}$, $\mathrm{C_I-C_N}$ and $\mathrm{C_I-C_{Ga}}$,
have also been considered. $\mathrm{C_N-C_{Ga}}$ is favorable in the
upper half of the band gap ($n$-type) region, whereas
$\mathrm{C_I-C_N}$ and $\mathrm{C_I-C_{Ga}}$ are favorable in the
lower half of the band gap ($p$-type) region. Finally, Complexes
with vacancies are also examined, in particular
$\mathrm{C_N-V_{Ga}}$ and $\mathrm{C_{Ga}-V_N}$.
The former is favorable with lower formation energy comparing to the latter
close the CBM, but is unstable as a complex in the $n$-type region with
negative value of binding energy.
The latter is favorable in the $p$-type region comparing to the former.

From the calculated formation energies we have evaluated the
thermodynamic transition levels. These are directly related to the
thermal activation energies observed in experimental techniques such
as DLTS. In addition, by calculating Franck-Condon shift, optical
activation energies, which can be obtained by optical techniques
such as DLOS, have been evaluated from the thermal activation
energies. We compare our calculated values of activation energies
with the energies of experimentally observed C-related trap levels.

Using the information on the transition levels we have assigned
the C-related trap levels, whose physical form was unknown before.
It should be noted that these assignments are performed based only
on the positions of the trap levels and their concentrations are not
taken into account.
The trap level observed at $E_c - 3/E_v + 0.9$\,eV is likely due to
the (0/$-$) level of $\mathrm{C_N}$
with possible contributions from $\mathrm{C_I}$ and $\mathrm{C_I-C_N}$.
Based on earlier investigations
performed employing LDA, two different energy levels: $E_c - 3.2$\,eV
level by MCTS (thermal method) and $E_c - 3.28$\,eV by DLOS
(optical method), were attributed to the (0/$-$) level of
$\mathrm{C_N}$. However, our HSE calculation shows that the origin
of the $E_c - 3.2$\,eV level is the (+/0) level of $\mathrm{C_N}$ and
of the level at $E_c - 3.28$\,eV is
the (+/0) of $\mathrm{C_N-V_{Ga}}$.

Based on LDA results, the trap level observed at $E_c - 1.35$\,eV
was unanimously attributed to the (0/2$-$) state of $\mathrm{C_I}$.
However, the outcome of our HSE calculation suggests its origin is
the (0/$-$) and/or (+/0) states of the $\mathrm{C_I-C_{Ga}}$ complex.
The trap at $E_c - 1.94/2.05$\,eV is likely due to two
configurations. One is the (2+/+) of $\mathrm{C_I-C_{Ga}}$ and the other
is the (2+/+) of $\mathrm{C_{Ga}-V_N}$.
Looking at the trap level located at $E_c - 2.7/2.8$\,eV we argue
that its origin stems from the (2+/+) state of the
$\mathrm{C_I}$ and the (3+/2+) state of the $\mathrm{C_I-C_{Ga}}$.

Multiple contributors are likely to be responsible for the trap at
$E_c - 0.4$\,eV. Specifically we find that the (0/$-$) of
$\mathrm{C_I}$
and the (+/0) of $\mathrm{C_{Ga}-V_N}$ have energies that are close to the measured
value.
Finally, the trap at $E_c - 0.1$\,eV was attributed to
$\mathrm{C_{Ga}}$ based on LDA results. However, based on our HSE
calculations, there are no gap states due to $\mathrm{C_{Ga}}$.
Therefore the origin of this trap level is still unclear, but the
most likely candidate is the (+/0) of $\mathrm{C_{Ga}-V_N}$ and/or the
(0/$-$) of $\mathrm{C_I}$. In this
case these levels have same origin as $E_c -0.40$\,eV level.

Our calculated LVM results would provide additional information for the experimental
detection of these carbon related impurities.

\begin{acknowledgments}

The authors thank K. Jones and R. Tompkins of
the Army Research Laboratory and T. Moustakas of Boston University
for many discussions and their help in
understanding the experimental techniques. The authors are grateful
to A. F. Wright, S. Lee and N. Modine from Sandia National
Laboratory and S. Sharifzadeh of Boston University for discussing
the results of our work.

The authors gratefully acknowledge financial support from the U. S.
Army Research Laboratory through the Collaborative Research Alliance
(CRA) Grant No.\ W911NF-12-2-0023 for MultiScale multidisciplinary Modeling of Electronic
materials (MSME)\@. This work was performed using DoD HPCMP
supercomputing resources and computational resources provided by the
2014 Army Research Office Grant No.\ W911NF-14-1-0432 DURIP Award made to E. Bellotti.
\end{acknowledgments}


\begin{thebibliography}{63}
\bibitem [{sha()}]{shallow}%
  \BibitemOpen
  \href@noop {} {}\bibinfo {note} {In this case we have used the term shallow
  acceptor/donor rather loosely. Obviously from the practical standpoint an
  activation energy larger than several thermal energies will lead to a small
  fraction of ionized donors and acceptors. Nevertheless, we will use the
  qualifiers shallow and deep to distinguish between states that are closer or
  farther from the CBM/VBM.}\BibitemShut {Stop}%
\bibitem [{4pl()}]{4plus}%
  \BibitemOpen
  \href@noop {} {}\bibinfo {note} {Here the $4+$ charge state was defined by
  removing four electrons from the whole system, not from the C atom. In fact,
  electron deficiency is localized around the C-N$_3$ (C and surrounding three
  N atoms) structure.}\BibitemShut {Stop}%
\bibitem [{hol()}]{holecapture}%
  \BibitemOpen
  \href@noop {} {}\bibinfo {note} {The hole capture process is not explicitly
  considered here because none of the experimental optical activation energies
  cited in this paper are obtained by the hole capture process.}\BibitemShut
  {Stop}%
\end{thebibliography}

%
%
\end{document}